\documentclass[twocolumn]{aastex63}
\usepackage{amsmath}
\usepackage{multirow}

\newcommand{\rev}[1]{{\color{black}{\textbf{#1}}}}

\received{2020/12/16}
\revised{2021/05/06}
\accepted{2021/05/19}
\submitjournal{ApJ}

\shorttitle{Fast Multipole Method for Collisional Dynamics}
\shortauthors{Mukherjee et al.}

\graphicspath{{./}{final_plots/}}

\begin{document}

\title{Fast Multipole Methods for $N$-body Simulations of Collisional Star Systems}

\correspondingauthor{Diptajyoti Mukherjee}
\email{diptajym@andrew.cmu.edu}

\author{Diptajyoti Mukherjee}
\affiliation{McWilliams Center for Cosmology, Department of Physics, Carnegie Mellon University, 
Pittsburgh, PA 15213, USA}

\author{Qirong Zhu}
\affiliation{McWilliams Center for Cosmology, Department of Physics, Carnegie Mellon University, 
Pittsburgh, PA 15213, USA}

\author{Hy Trac}
\affiliation{McWilliams Center for Cosmology, Department of Physics, Carnegie Mellon University, 
Pittsburgh, PA 15213, USA}
\affiliation{NSF AI Planning Institute for Physics of the Future, Carnegie Mellon University, Pittsburgh, PA 15213, USA}

\author{Carl L.~Rodriguez}
\affiliation{McWilliams Center for Cosmology, Department of Physics, Carnegie Mellon University, 
Pittsburgh, PA 15213, USA}



\begin{abstract}

Direct $N$-body simulations of star clusters are accurate but expensive, largely due to the numerous $\mathcal{O} (N^2)$ pairwise force calculations. To solve the post-million-body problem, it will be necessary to use approximate force solvers, such as tree codes. In this work, we adapt a tree-based, optimized Fast Multipole Method (FMM) to the collisional $N$-body problem. The use of a rotation-accelerated translation operator and an error-controlled cell opening criterion leads to a code that can be tuned to arbitrary accuracy. We demonstrate that our code, {\tt\string Taichi}, can be as accurate as direct summation when $N> 10^4$. This opens up the possibility of performing large-$N$, star-by-star simulations of massive stellar clusters, and would permit large parameter space studies that would require years with the current generation of direct summation codes. Using a series of tests and idealized models, we show that {\tt\string Taichi} can accurately model collisional effects, such as dynamical friction and the core-collapse time of idealized clusters, producing results in strong agreement with benchmarks from other collisional codes such as {\tt\string NBODY6++GPU} or {\tt\string PeTar}.  Parallelized using {\tt\string OpenMP} and {\tt\string AVX}, {\tt\string Taichi} is demonstrated to be more efficient than other CPU-based direct $N$-body codes for simulating large systems.  With future improvements to the handling of close encounters and binary evolution, we clearly demonstrate the potential of an optimized FMM for the modelling of collisional stellar systems, opening the door to accurate simulations of massive globular clusters, super star clusters, and even galactic nuclei.

\end{abstract}


\section{Introduction} \label{sec:intro}

The collisional $N$-body problem, in which the gravitational dynamics of $N$ particles in a system are modeled over time, is one of the most challenging problems in modern computational physics.  The stellar environments represented by such models, such as open, globular, and nuclear star clusters, contain some of the highest known densities of stars and compact objects and can produce many interesting astrophysical systems and transients.  Systems such as X-ray binaries \citep[e.g.,][]{Clark1975,Davies1998,Ivanova2008,Hailey2018}, recycled millisecond pulsars \citep[e.g.,][]{Rappaport1989,Kulkarni1990,Sigurdsson1995,Ye2019}, cataclysmic variables \citep[e.g.,][]{Ivanova2006, Pooley2006}, and merging binary black holes \citep[e.g.,][]{PortegiesZwart2000,Rodriguez2015} can be produced with orders of magnitude more efficiency through dynamical encounters in dense star clusters than through typical stellar evolutionary processes. The compact objects within globular clusters are believed to be the sources of gravitational waves detected by LIGO \citep[e.g.,][]{abbott2020gw190814, abbott2020gw190425, abbott2020gw190412, abbott2017gw170814}.  Collisions of stars and compact objects in the central region are thought to be responsible for the formation of intermediate-mass black holes \citep[e.g.,][]{Freitag2006,Gurkan2006,Giersz2015}, and possibly even the seeds of supermassive black holes at high redshift \citep[e.g.,][]{ebisuzaki2001missing}. These black hole seeds grow along with their host galaxies and will be the targets of incoming space detectors such as LISA \citep{amaro2017laser} and Tianqin \citep{luo2016tianqin}. 

To circumvent the difficulties associated with a direct integration approach to the $N$-body problem, approximate techniques such as H\'enon-style Monte Carlo approaches \citep[e.g.,][]{Henon1971,Henon1971a,Giersz1999,Joshi1999,pattabiraman2013parallel, rodriguez2016million, hypki2017mocca}, or approximate solvers of the collisional Fokker-Planck (FP) equation \citep[e.g.,][]{vasiliev2017new}, are often used to follow the dynamics.  The foundation of both approaches is a statistical treatment of uncorrelated two-body encounters over long time \citep{chandrasekhar1942principles}. For these methods to work,  the classic Chandrasekhar's formulae for dynamical friction
(first-order) and diffusion (second-order) coefficients are derived under somewhat
strong assumptions, which neglects coherent motions of each individual particles, i.e.,
resonances \citep{Meiron2019} or self-gravity \citep{Lau2019}. Moreover, the
uncertainties in the Coulomb logarithm need to be calibrated against the numerical
method \citep{Merritbook}. Therefore, the most versatile and adaptive method is 
direct integration of $N$-body system. However, direct $N$-body modeling
is notoriously expensive to run and difficult to interpret \citep[e.g.,][]{Miller1964}.

With the increase in computational power, direct $N$-body methods have become more accessible for performing larger simulations with $N \sim 10^6$. This makes them one of the preferred means to simulate star clusters today. Although accurate, the simulations are computationally expensive owing to the $\mathcal O (N^2)$  complexity of the direct forces calculation algorithm. For example, the DRAGON simulations \citep{wang2016dragon} took $\sim$ 8600 hours using 160 {\tt\string Xeon-2560} cores and 16 {\tt\string K20m} GPUs to simulate a globular cluster containing $10^{6}$ stars. Thus, solving the post-million-body problem using direct $N$-body codes presents a considerable challenge and is of interest to the astrophysical community.

One method that has been in widespread use is the differential treatment of long-range and short-range interactions. The Ahmad-Cohen scheme \citep{ahmad1973numerical}, which has been adopted in {\tt\string NBODY6} \citep{aarseth2003gravitational}, allows the long-range interactions to be updated less frequently compared to the short-range interactions, thereby requiring fewer force calculations per average time. 
The alternatives to the expensive direct summation method are approximate force solvers. Among those, the  Barnes-Hut (BH) tree \citep{barnes1986hierarchical} 
is a natural candidate. Tree codes use hierarchical decomposition and multipole expansions to calculate gravitational forces between the particles. The latter results in an algorithmic complexity of $\mathcal O (N \log (N))$, which is a major improvement from that of direct summation methods. Although tree codes have found widespread usage in collisionless dynamics, they have found limited usage for collisional dynamics simulations so far \citep[e.g.,][]{McMillan1993, aarseth1999nbody1}, the primary reason being the concern of force accuracy. \cite{iwasawa2015gpu} indicate that one of the other reasons may lie in the the fact that collisional simulations adopt individual or block time stepping. This would decrease efficiency since the particle tree would have to be reconstructed every time step. Despite this issue, ${\rm P^{3}T}$ (particle-particle particle-tree) codes which combine the force splitting with the Barnes-Hut tree have come out recently \citep[e.g.,][]{iwasawa2015gpu, wang2020petar}. \cite{wang2020petar} have demonstrated that their code is highly competitive 
compared to the direct summation code.

BH tree codes are not the only option here.
\cite{dehnen2014fast} describes a more efficient approach compared to the traditional tree code by adapting the Fast Multipole Method \citep[(FMM: )][]{greengard1987fast, cheng1999fast} to collisional dynamics. \cite{dehnen2014fast} presents various optimizations to the traditional FMM algorithm, which makes it suitable for adoption into star cluster simulations. An algorithmic complexity of $\mathcal O (N)$ and even sub-$N$ complexity for special cases is demonstrated. The efficiency and the well-behaved error properties of this approach make it an attractive alternative to the tree
codes. However, adaptation of FMM for collisional dynamics is missing. FMM has been used for collisionless $N$-body simulations before \citep[e.g.][]{dehnen2000very, dehnen2002hierarchical}.  Collisional simulations demanding higher accuracies and proper treatment of individual particles complicated the adoption of FMM.

An interesting question arises about whether a full-fledged $N$-body code using FMM could be as accurate and efficient as direct summation codes. The issue with accuracy echoes same concern voiced by \cite{Press1986}. In this study, we adopt the FMM algorithm presented in \cite{dehnen2014fast} into our code, {\tt\string Taichi} \citep{zhu2020momentum}, to perform various tests in collisional dynamics.

In Section \ref{sec:fmm} we describe briefly the modifications that were made to \cite{zhu2020momentum} to adapt {\tt\string Taichi} to collisional simulations. In Section \ref{sec:methods} we describe the tests performed with the code. Section \ref{sec:results} goes into the results of the various tests and discusses the significance of the results. It is followed by discussion in Section \ref{sec:discussion}, future work in  Section \ref{sec:future_work}, and conclusions in Section \ref{sec:conclusion}.

\section{Taichi for Collisional Dynamics} 
\label{sec:fmm}

\begin{figure*}
    \includegraphics[width=0.9\textwidth]{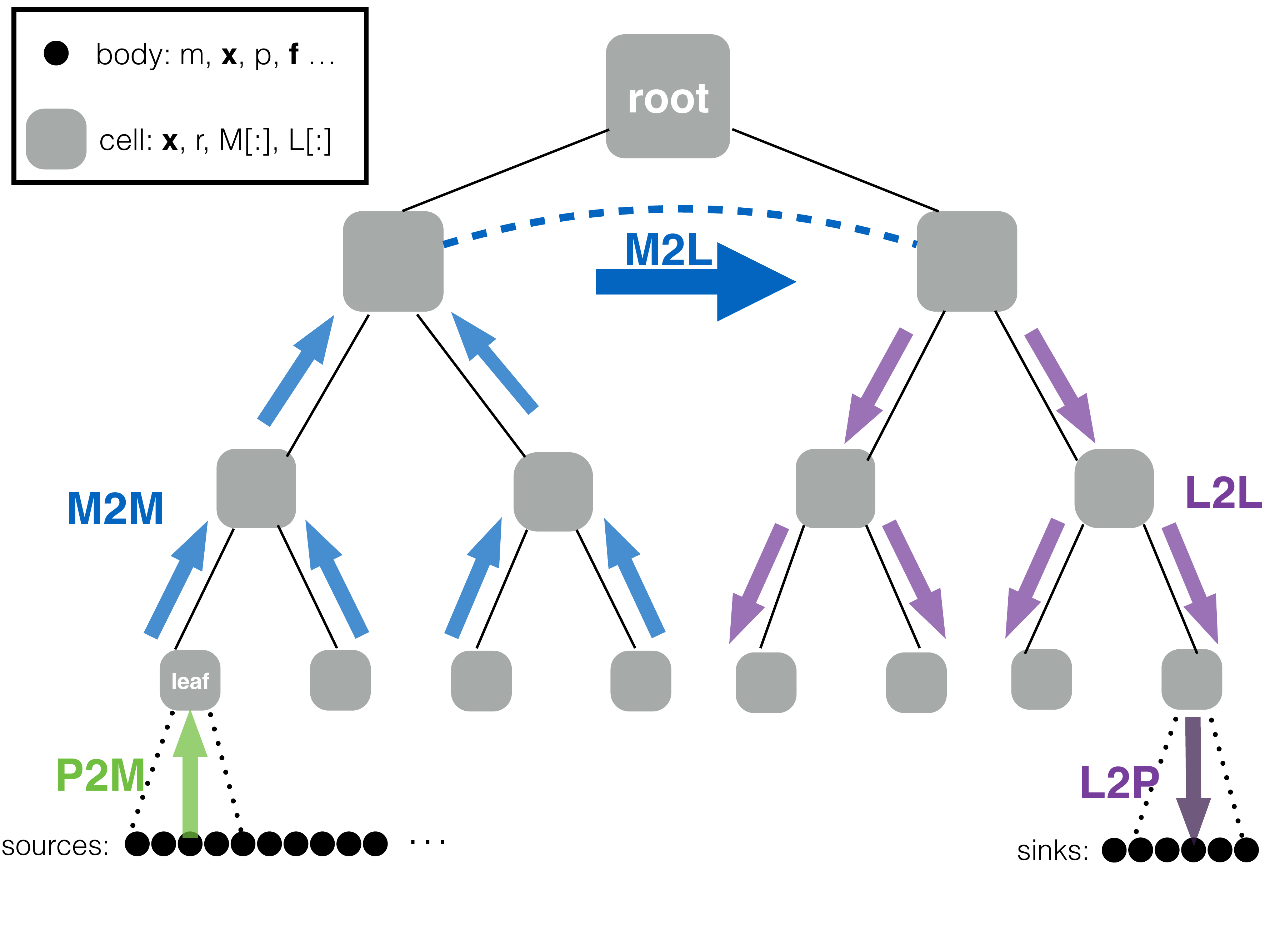}
    \caption{Flowchart \citep[adapted from][]{yokota2012tuned} illustrating various working parts of the FMM algorithm. For information on the abbreviations, check out section \ref{subsec:fmm_desc}. The particles are arranged in an oct-tree, which provides a complete and hierarchical description of a set of particles, and will be ready for calculating the far-field of any cell consisting of a group of particles. The individual bodies contain information regarding the mass, position, velocity, and force experienced. The cells contain information regarding the position, radius, and multipole and local expansions. For clarity, we show the left half of the tree, consisting of only sources, and the right half, consisting of only sinks. The interaction list is obtained by a dual-tree walk \citep{dehnen2002hierarchical} to determine which pair of cells can use approximations.    }
    \label{fig:flowchart}
\end{figure*}

As emphasized in the introduction, an efficient FMM algorithm is the
core of our study. \cite{zhu2020momentum} implements a collisionless $N$-body with FMM and individual time steps into a code called {\tt\string Taichi}. In this work, we build on \cite{zhu2020momentum} and make important modifications to enable it to be used for collisional simulations. The modifications are made in three different areas: usage of solid spherical harmonics instead of Cartesian coordinates for the multipole expansion, inclusion of an explicit error control algorithm, and, lastly, usage of rotation operators to speed up multipole expansions. In the following subsections we describe the overall structure of FMM and the modifications over \cite{zhu2020momentum} in more detail.

\subsection{Dual-tree Walk}
Figure \ref{fig:flowchart} illustrates the flow of FMM adopted as a force solver in {\tt\string Taichi}. The particles are hierarchically arrange in an oct-tree. Each leaf cell first collects the multipole expansion with a Particle-to-Multipole (P2M) kernel. The multipoles are then passed recursively upward by the parent nodes until the root using a Multipole-to-Multipole (M2M) kernel. At this point, we have a complete description of all the cells for their far-field gravity if necessary.

Next, we pass the root to a dual-tree walk \citep{dehnen2000very} and determine which interactions can be approximated. Our implementation of FMM uses a task parallel version of the dual-tree walk. Approximated gravitational force between the cells is calculated with a Multipole-to-Local (M2L) kernel for well-separated pairs of cells. The local expansions are then passed recursively down the nodes to the leaf cells. This step is achieved with a Local-to-Local (L2L) kernel. Finally, the force and the potential energy of each particle are determined based on the local expansions of the leaf cells they reside in using a Local-to-Particle (L2P) kernel. To calculate the near-field contribution to the leaf cells, which cannot be handled using multipole expansions, a direct summation is carried out instead.

\subsection{FMM with Solid Spherical Harmonics}
We use the same architecture in \cite{zhu2020momentum} for 
tree building, dual-tree walk, and direct summation.
The FMM implementation in \cite{zhu2020momentum} is 
based on a Taylor expansion in Cartesian coordinates
\citep{dehnen2000very}, 
which is sufficient for collisionless systems where
a modest force accuracy is needed. In
collisional systems, a more stringent force accuracies are 
presumably necessary to follow changes
to the particle's orbit over many dynamical times. While
previous studies based on the BH tree method indicate that
quadrupole expansion with opening angle $\theta=0.5$ is 
sufficient, a systematic study of force accuracies on the collisional
dynamics is desired. To be conservative, we consider that the 
approximate force aiming at the round-off level of single-precision 
floating point arithmetic should be safe. To that end, we adopt 
the approach using \textit{solid harmonics} outlined in 
\cite{dehnen2014fast}.

For $1/r$ interaction, the solid harmonics essentially 
form a complete and orthogonal basis for all the 
multipole moments and their far-field potentials. Up
to order $p=2$, the explicit expressions of these
functions \citep[table 3 in][]{dehnen2014fast} read
\begin{equation}
1, x, y, z, 3(x^2-y^2), 6xy, 3xz, 3yz, 3z^2-r^2;
\end{equation}
One can see that these are the traceless part of the 
the Cartesian expressions \citep[see also][]{Coles2020}. 
We follow Equations (52), (53), and (54) in \cite{dehnen2014fast}
to generate the regular ($\Upsilon_{n}^{m}$) and
irregular (${\Theta_{n}^{m}}$) parts of solid spherical harmonics. The former is used in the multipole moments 
\begin{equation}
\mathcal{M}_{n}^{m} = \sum_i m_i \Upsilon_{n}^{m} (\mathbf{r}_i),
\label{eq:moments}
\end{equation}
where the summation is taken over each particle $i$
with mass $m_i$ and displacement vector $\mathbf{r}_i \equiv (x_i, y_i, z_i)$.
The latter is used in the expansion and translation operations.
For more detailed mathematical properties of $\Upsilon_{n}^{m}$
and ${\Theta_{n}^{m}}$ and their translations, 
we refer the reader to \cite{dehnen2014fast}.
All the factorials present in the normalization
coefficients are precomputed in a look-up table. The adoption
of the solid harmonics enables us to truncate the expansion to 
a very high order, $p\gg 10$, which would otherwise become 
cumbersome for Cartesian expansions.

\subsection{Error-controlled multipole-acceptance criteria}
\label{subsec:fmm_desc}

The second important modification is an error-controlled
multipole-acceptance criterion (MAC) over the conventional varying $\theta$ as in \cite{zhu2020momentum}. 
We aim at some fractional force error for the force calculation:
therefore, we adopt a scalar force as a quick but reasonable 
approximation of the actual accelerations according to 
\begin{equation}
    f_i = \sum_{i\ne j} \frac{G m_j}{\left|\mathbf{x}_i - \mathbf{x}_j \right|^2}.
\end{equation}

Before the actual dual-tree walk, we first estimate
the total scalar force $f$ with $\theta < 1$. Each 
cell collects the contribution from far-field cells
as $\sum \frac{M_c} {r^2}$, where ${M_c}$ is the cell 
mass and $r$ the separation between the cells. 
Next, the actual force 
calculation proceeds with the following 
MAC:
\begin{equation}
\theta < 1\ \land E_{A\rightarrow B} \frac{M_A}{r^2} < \epsilon f_{B} \land E_{B\rightarrow A} \frac{M_B}{r^2} < \epsilon f_{A}
\end{equation}
where $f_{A}$ and $f_{B}$ are the minimum of scalar force $f$ 
for those particles in cells $A$ and $B$ respectively. The tolerance parameter $\mathbf{\epsilon}$ directly 
controls the final force accuracies by FMM.
We note that the above criterion which slightly differs
from \cite{dehnen2014fast}, additionally ensures that the 
forces among particles are symmetric at some extra cost.

The error coefficients $E_{A\rightarrow B}$ 
and $E_{B\rightarrow A}$ are entirely
determined by the multipole moments of $A$, $B$ 
themselves. With $\mathcal{P}_n$ defined as the sum of all the $(2n+1)$ 
multipole terms on order $n$ as
\begin{equation}
\mathcal{P}^2_n  = \sum_{m=-n}^{n}(n-m)!(n+m)! \left| \mathcal{M}_n^m \right| ^2, 
\end{equation}
where $M_n^m$ are the multipole moments of the cell as in Eq.~(\ref{eq:moments}), 
and 
\begin{equation}
E_{A\rightarrow B} = \frac{8 \max\{r_A, r_B\}}{r_{A}+r_{B}} 
 \frac{1}{M_{A}} \sum_{k=0}^{p} \frac{p!}{k!(p-k)!}  \frac{\mathcal{P}_{k,A}r_{B}^{p-k}} {r_{}^p}, 
\end{equation}
 where $M_A$ is the total mass of source cell $A$, i.e., its $ M_{0}^{0}$ moment, $r_A$ and $r_B$ are the sizes of each cell, and $r_{}$ is their separation.

\subsection{Rotation-accelerated M2L Operations}
The last, but not the least, improvement is a fast M2L kernel.
To speed up these expensive operations, we adopt the rotation-accelerated $\mathcal{O}(p^3)$ approach \citep{cheng1999fast, dehnen2014fast}. Additionally, we generate and save the swapping matrices for expansion order $p\le30$.  These swapping matrices are essential for M2L translation operations, where the new $z$-axis is aligned with the interaction direction as follows: the multipole moments of the source cell are rotated in the $z$-direction, then with its $x$ and $z$ coordinates swapped, rotated in $z$-direction again, and with its $x$ and $z$ swapped again. The translation in the new coordinates features$\mathcal{O}(p^3)$ complexity instead of $\mathcal{O}(p^4)$.

We use {\tt\string OpenMP} to speed up the dual-tree walk using
task model with atomic clause used
to update of multipole moments \citep{Fortin2019}. The near-field contribution is handled by 
a direct summation kernel, which is vectorized 
using {\tt\string AVX} intrinsics as in \cite{zhu2020momentum}. 
Time integration closely follows the {\tt\string HUAYNO} code \citep{Pelupessy2012} with little changes. A brief overview of the integrator and the time-symmetric time-stepping method used in {\tt\string Taichi} is provided in Appendix A.

\section{Tests} \label{sec:methods}

To measure how effective FMM is at simulating star clusters, we compare models of idealized clusters to the results of other codes and theoretical predictions. We use a homogeneous Plummer model \citep{plummer1911problem} to generate our initial conditions. The initial conditions are generated using the tool {\tt\string MCLUSTER} \citep{kupper2011mass}. Each test involves a number of independent realizations, and the results are derived after taking statistical averages over these realizations. For comparison, {\tt\string NBODY6++GPU} \citep{wang2015nbody6++} without GPU acceleration enabled is used to perform the direct $N$-body simulations. The same tests are also performed using the direct version of {\tt\string Taichi} to ensure that the presence of the second-order {\tt\string HOLD} integrator did not bias the results in any manner. In addition, Fokker-Planck simulations are performed using {\tt\string Phaseflow} \citep{vasiliev2017new} to compare density profiles over the course of evolution until core collapse. For the dynamical friction tests, {\tt\string PeTar} \citep{wang2020petar} is used as a benchmark along with the other codes mentioned. In all of the tests Hénon \citep{heggie1986standardised} units are used. In these units, $G=M=1$ and $E_0 = -0.25$, where $M$ is the total mass and $E_0$ is the initial total energy. A summary of the input parameters used for the simulations has been provided in Table 1. 

All tests are done on 28-core {\tt\string Intel Xeon E5-2635 v3} nodes. {\tt\string Taichi}, {\tt\string NBODY6++GPU}, and {\tt\string PeTar} are run with only {\tt\string OpenMP} and {\tt\string AVX} enabled.
\subsection{Tests Performed} \label{subsec:tests}
Three different tests are performed in order to examine how FMM compares to direct $N$-body codes. In the first set of tests, we compare how accurate the FMM algorithm is compared to the direct summation method by examining force discrepancies. In the second set of tests, we evolve idealized Plummer models until core collapse to measure global properties, including conservation of energy, evolution of Lagrangian radii, core radii, and density function. Finally, we compare dynamical friction effects via the inspiral of a massive particle in a field of smaller mass stars. We also perform scaling tests to examine how {\tt\string Taichi} scales with the number of cores within a node. The tests are presented in more detail in the upcoming sections. 

\subsubsection{Force Accuracies}

 We compare how the forces on individual particles vary between the direct and FMM versions of {\tt\string Taichi} after a single time step. We construct a Plummer sphere with $10^5$ particles using {\tt\string MCLUSTER} and integrate it using both direct and FMM versions of {\tt\string Taichi} with different input accuracies and multipole expansions. The relative force accuracies are then computed using the L2 norm as follows: 
\begin{gather}
    \mathbf{\delta f} = \mathbf{f_{\mathrm{direct}}} - \mathbf{f_{\mathrm{fmm}}} \\
    \frac{\delta f}{f} = \frac{\left| \mathbf{\delta f} \right|}{\left|  \mathbf{f_{\mathrm{direct}}}  \right|}
\end{gather}
We look at how the distribution of relative force accuracies varies with changing input parameters. Using the grid of simulations performed, we construct a heat map plotting the median and 99.99th percentile fractional force accuracies and histograms plotting the distribution of relative force accuracies. In addition, we look at the time taken by the FMM simulations with different input accuracy and multipole parameters to construct heat maps showing the variation of the integration time and the Poisson step time with the change in input parameters.

\subsubsection{Core Collapse of a Plummer sphere}
{\tt\string MCLUSTER} is used to generate three sets of 16 simulations each containing $N=$ 1024, 2048, and 4096 particles. The clusters are evolved until core collapse ($\sim 15 t_{\mathrm{rh}}$) where $t_{\mathrm{rh}}$ is the half-mass relaxation time, which for the Plummer model is defined as
\begin{equation}\label{t_rh}
    t_{\mathrm{rh}} = 0.206  \frac{ a^{3/2} N}{\log(\gamma N)}
\end{equation}
where $a$ is the characteristic scale or Plummer radius and ${\log(\gamma N)}$ refers to the Coulomb logarithm. All of the models start in virial equilibrium and do not contain any primordial binaries.

Post-collapse treatment is unfeasible since {\tt\string Taichi} does not include regularization treatment for hard binaries, and as such the simulations take a long time to finish post-collapse. {\tt\string Phaseflow} is used to simulate Fokker-Planck models of the clusters. The scale radius of the cluster is set to $0.59$ which corresponds to Hénon units. The Coulomb logarithm ($\log \Lambda \equiv \gamma N$) is calculated by setting $\gamma=0.11$. This value is representative of clusters with a uniform mass function \citep{giersz1994statistics}. 

\subsubsection{Dynamical Friction}
In order to measure whether FMM can accurately model dynamical friction, we seek to reproduce the black hole inspiral test performed by \cite{rodriguez2018new}. In this test, a massive particle several times the mass of the stars in the cluster is introduced on a circular orbit at the viral radius of the cluster. Its position relative to the center of mass of the cluster is tracked. The time taken by the massive particle to inspiral to the center of the cluster can be modeled analytically. For a massive particle with mass $m$ starting at a radius $r$ with a circular orbit in a Plummer model, \cite{rodriguez2018new} provide the rate of change of $r$ as
\begin{equation}\label{dr_dt}
    \frac{dr}{dt} = \frac{-8\pi G^{2} \log \Lambda \chi m r} {V_{c} ^{3} \left[1+3(1+\frac {r^{2}}{a^{2}})^{-1}\right]}
\end{equation}
where $\log \Lambda$ is the Coulomb logarithm, $m$ is the mass of the massive object, $V_{c}$ is the circular velocity of the massive object at a distance $r$ and $a$ is the scale radius of the cluster. $\chi \equiv {\rm erf}(X) - 2X{\rm exp}(-X^2)/\sqrt{\pi}$, $X \equiv V_{c}/(\sqrt{2}\sigma (r))$, where $\sigma (r)$ is the velocity dispersion at a radius $r$ has been used in equation \ref{dr_dt}.  In Hénon units, $G=1$ and $a=0.590$.  $\gamma$ is set to be $0.01$ in this equation \citep{rodriguez2018new}. These tests can also be used to determine $\gamma$ since the analytic solution is very sensitive to the value of $\gamma$.
More details on the derivation of this equation can be found in \cite{rodriguez2018new} and \cite[][chapter 8]{binney2011galactic}. 

Using {\tt\string MCLUSTER}, 30 independent realizations containing $10^{4}$ stars are generated. The last star in the initial conditions is then replaced with either an object either 10 times more massive or 20 times more massive depending on the simulation. The massive object is placed one virial radius from the center of mass of the cluster on a circular orbit with velocity
\begin{equation}
    V_c(r_{\rm vir}) = \sqrt{\frac{GM(r_{\rm vir})}{r_{\rm vir}}} ,
\end{equation}
where $M(r_{\rm vir})$ is the mass contained within one virial radius. Note that $V_c(r_{\rm vir}) \approx 0.799$ when $r_{\rm vir} = 1$.

\begin{table}[h!]

\begin{center}
\begin{tabular}{ |c|c|c| }

\hline
Code & Input Parameters \\
\hline
\multirow{3}{5em}{{\tt\string Taichi} FMM} & $\epsilon=10^{-7}$  \\ 
& $p=20$ \\ 
& $\eta=0.025$ \\ 
\hline
\multirow{2}{5em}{{\tt\string NBODY6}} & {\tt\string NNBOPT}$= 500$  \\ 
& $\eta_{\rm I} = \eta_{\rm R} = 0.01$ \\ 
\hline
\multirow{3}{5em}{{\tt\string PeTar}} & $\theta=0.3$  \\ 
& $\eta_{\rm Hermite}=0.1$ \\
& $\Delta t_{\rm tree} = $ Default \\
\hline
\end{tabular} 
\end{center}
\caption{Summary of the input parameters used for the different codes for the tests mentioned above. Note that for the direct version of {\tt\string Taichi} the same $\eta$ was used. The input parameters for {\tt\string NBODY6} were chosen in order to maximize the accuracy. In case of {\tt\string PeTar} the tree timestep parameter is calculated automatically by the code from the changeover radius. Please note that these input parameters were also used for the simulations performed in the Appendix.} 
\label{table:input_par_summary}

\end{table}

\section{Results} \label{sec:results}

\subsection{Accuracy} \label{subsec:accuracy}

\begin{figure*}[t]
\includegraphics[width=1.0\textwidth]{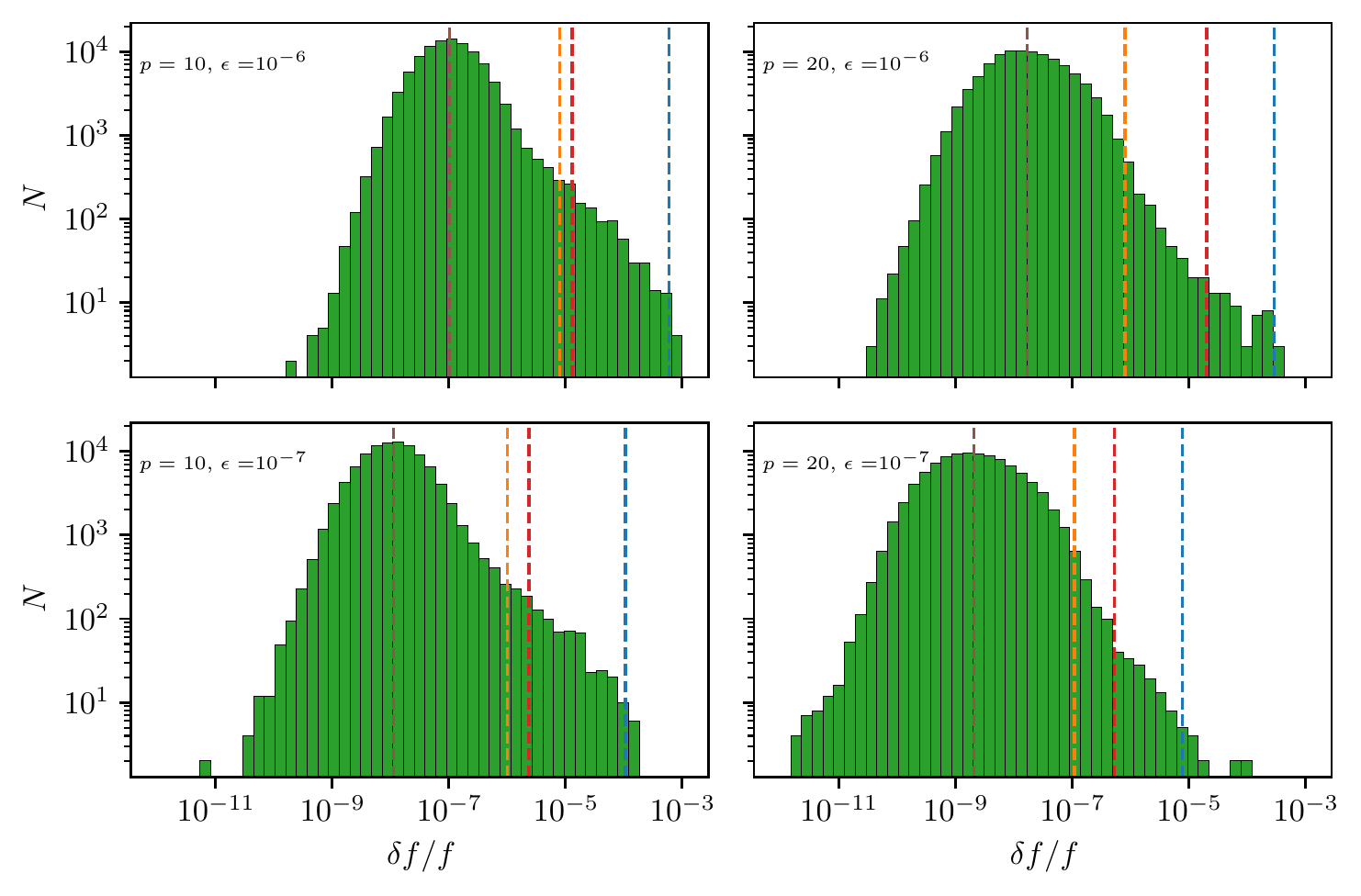}
\caption{A  histogram of the relative force error ($\delta f / f$). The particles are binned by their relative force errors. The brown line represents the median value, the yellow line represents the 99th percentile value, and the red line represents the rms value, and the blue line represents the 99.99th percentile values of relative force error. Increasing the multipole order shifts the distribution to the left and reduces outliers.}
\label{fig:acc_hist}
\end{figure*}

\begin{figure*}[t]
\includegraphics[width=1.0\textwidth]{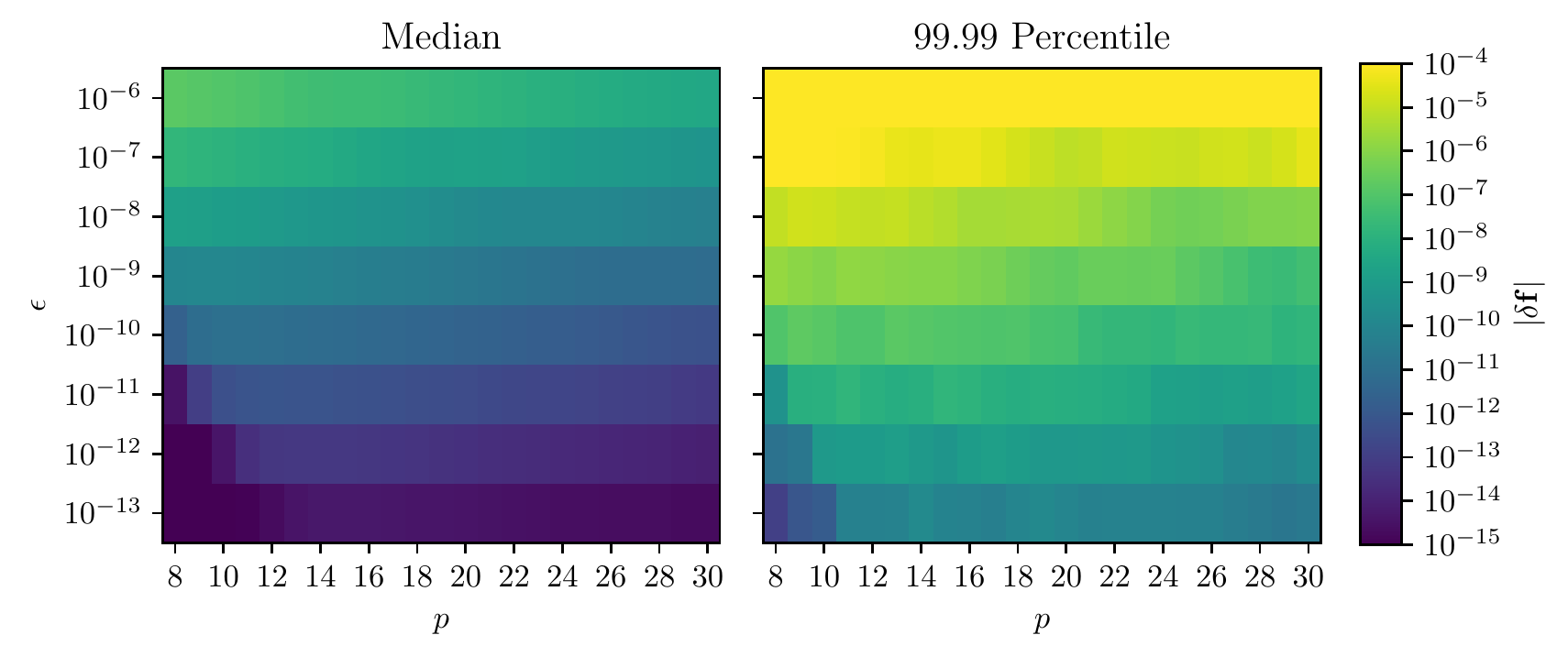}            
\caption{The relative force error distributions  as functions of input accuracy ($\epsilon$) and mulipole parameter ($p$). Left: the median relative force error is presented in this heat map. It is evident that the median fractional force error is extremely tightly controlled. In fact, for a given input accuracy, the median error is several orders of magnitude lower than it. Right: the 99.99th percentile fractional force error heat map is presented here. Since this value is representative of the number of outliers, we notice that the brighter patches indicate that the distributions contain more outliers than the darker portions. Within each row, there appears to be a fixed $p$ for which the 99.99th percentile values are lowest.} 
\label{fig:accuracy}
\end{figure*}

The global energy error does not provide a full picture of the validity of the simulations. As \cite{dehnen2014fast} mentions, even though in FMM the energy error is indicative of the average force errors, it can cloak individual force errors that might be large enough to question the validity of the simulations. Therefore, it is imperative that we use another means to measure the validity of the simulations. 

To determine the quality of force calculations, we look at the distribution of relative force errors for individual particles. In figure \ref{fig:acc_hist}, we compare the distribution of relative force errors after one time step while varying the input accuracy and the multipole expansion parameters. We find that the median of the distribution is always better than the input accuracy and improves as we increase the multipole expansion parameter. We also notice that the 99th percentile values of the distribution improve after increasing the multipole expansion parameter. In fact, the 99th percentile value is almost exactly equal to $10^{-7}$ when $\epsilon = 10^{-7}$ and $p=20$. The right tails of the the distribution plots show that the force errors can sometimes go as high up as $10^{-3}$. However, further analysis shows that the particles at the high-error  tail of the distribution are located predominantly at large radii. For the particles at large radii, the magnitudes of the forces are small, which can additionally lead to misleading large fractional force errors \citep{dehnen2014fast}. Also noted by \cite{dehnen2014fast}, few of the particles at the high-error tail could also lie at the center of the cluster, where the forces mostly cancel out, leading to a small force that contributes to a large fractional error.

We conclude that the input parameters can be tuned in order to reduce the force discrepancy between the direct summation  and the approximate values. To better understand how the parameters affect the distribution, we construct heat maps showing the distribution of the median and 99.99th percentile force error values since this gives us information about the distribution itself. Discrepancies between the two values indicate the length of the right tail of the error distributions, giving us an idea about the outliers. From the median value heat map in Figure \ref{fig:accuracy} we find that force error is well controlled by the the input accuracy parameter itself. Within any particular row, we see that increasing the multipole expansion parameter increases the overall force accuracy. The exception to this rule is seen in the lower left corner, where we have a combination of lower values of $\epsilon$ and $p$. For very low values of $\epsilon \leq 10^{-13}$, more cells are opened at low values of $p$ rather than at relatively larger values of $p$ because of the error estimation algorithm. In such cases, a number of cells can contain at most one particle, essentially reducing FMM to direct summation. This increases the relative force accuracy compared to direct summation but also results in a lot more pairwise force calculations, reducing efficiency. Thus, we find that the overall relative force accuracy decreases when we move from lower to higher $p$ values.

\subsection{Core Collapse of a Plummer Sphere} \label{subsec:res_global}

\begin{figure*}
    \includegraphics[width=1.0\textwidth]{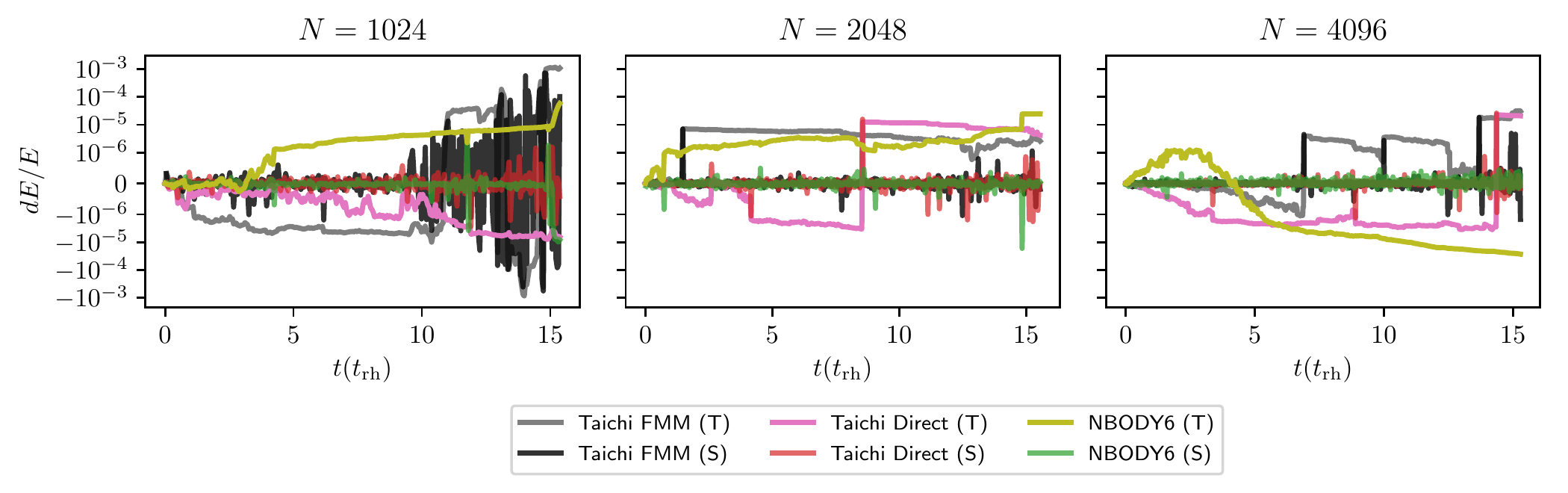}
    \caption{The relative energy errors represented as a function of the relaxation time. The curves with (T) denote the cumulative relative energy errors,m whereas the curves with (S) in them denote the relative energy error over one $N$-body time unit. The cumulative energy error starts growing more rapidly toward the end owing to the formation of hard binaries as the simulations approach core collapse. }
    \label{fig:energy_conservation}
\end{figure*}

\begin{figure*}[ht!]
\gridline{\fig{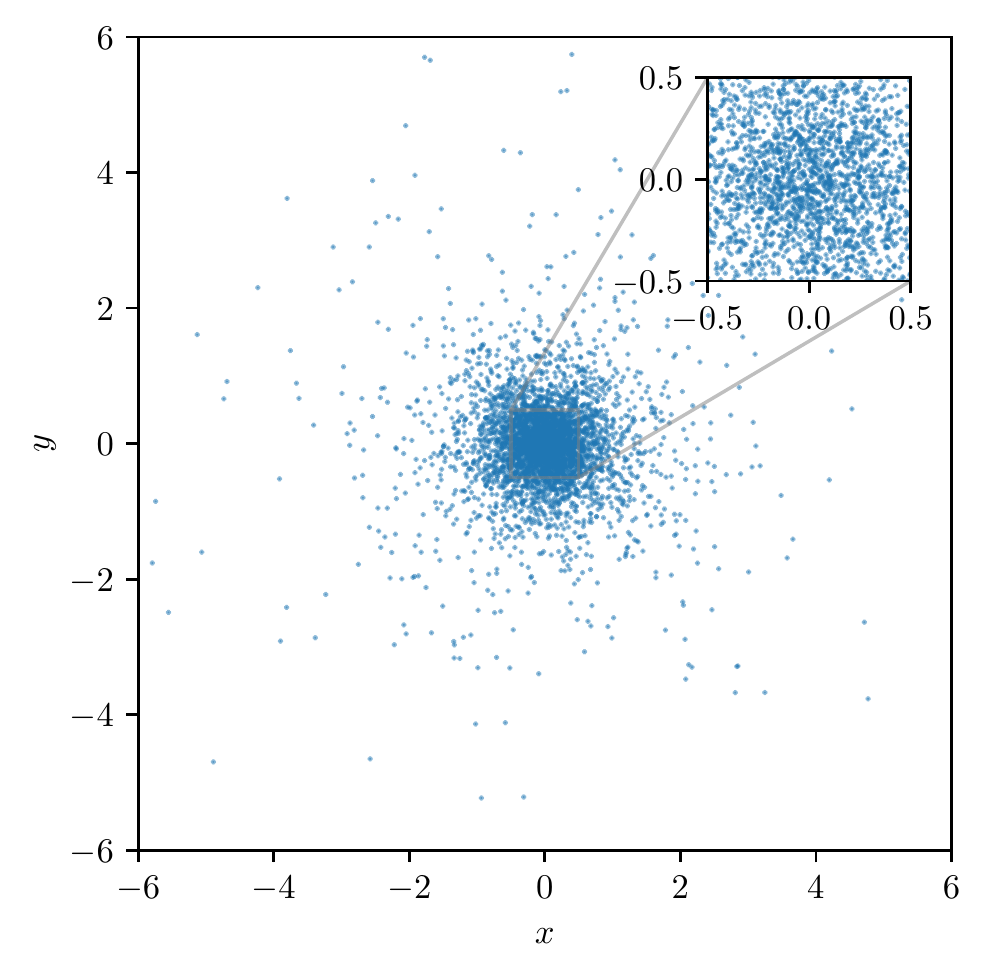}{0.50\textwidth}{}
            \fig{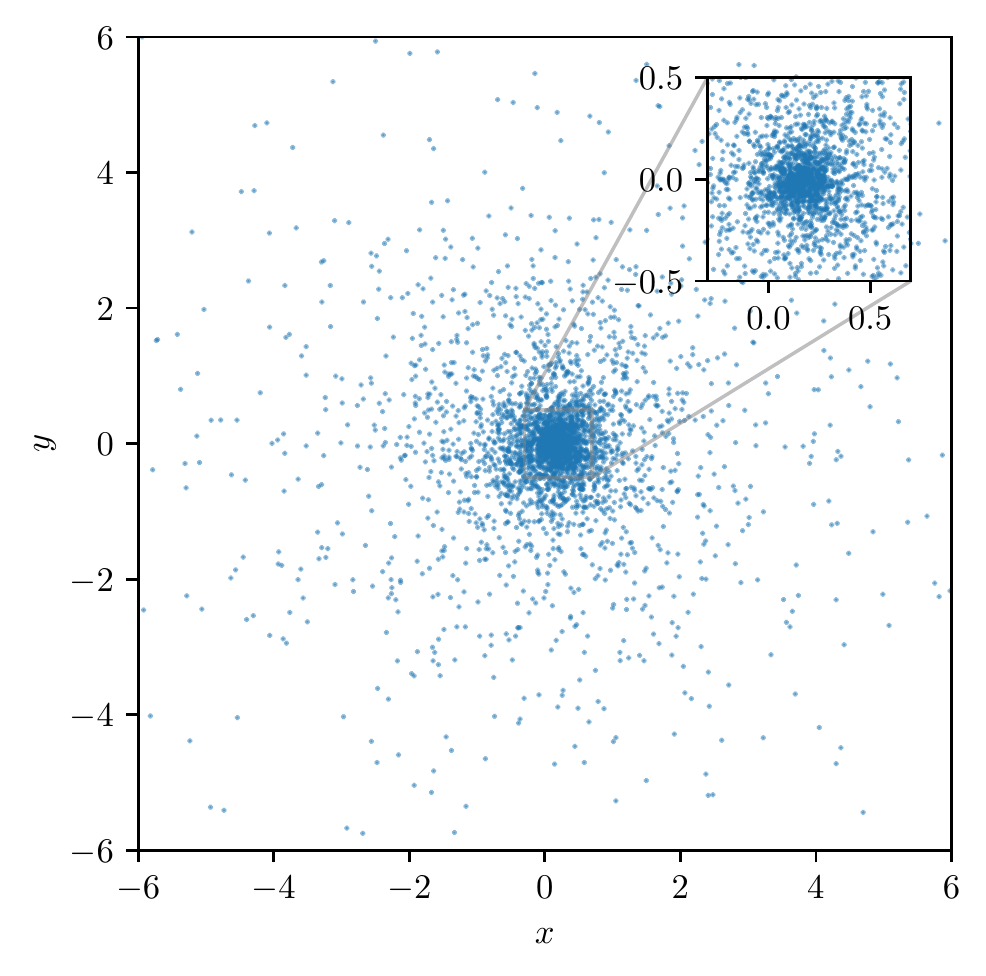}{0.50\textwidth}{}
            }

\caption{A cross-sectional scatter plot of an $N=4096$ particle simulation run with {\tt\string Taichi} FMM. Left: the particles at the initial timestep. The zoomed in area shows the region near the center of the cluster. Right: the particles right before core-collapse. One can clearly see the core that has been formed.}
\label{fig:visual}
\end{figure*}

As a preliminary check, we analyze the growth of energy errors over the long-term evolution of the system as shown in Figure \ref{fig:energy_conservation}. We notice that the growth of energy errors is relatively slow toward the beginning of the simulation and increases more rapidly as the cluster approaches core collapse. For {\tt\string Taichi} this is more evident since the usage of a symmetrized time steps ensures that the energy error grows more slowly in the beginning \citep{Pelupessy2012}. However, due to the lack of regularization, close encounters or few body interactions can lead to jumps in the energy errors, especially close to core collapse. We present an analysis of energy growth with softening enabled in Appendix B.

It should be noted that in the case unsoftened version of {\tt\string NBODY6++GPU}, the energy jumps are caused due to improper KS regularization switching as observed and noted by \cite{wang2020petar}. The cumulative energy conservation in the long term for the unsoftened version is typically $\mathcal{O} (10^{-4})$ for both {\tt\string Taichi} and {\tt\string NBODY6++GPU}. We notice that this is true for both direct summation and FMM versions of {\tt\string Taichi}. For the softened version, we find that the energy conservation in the {\tt\string Taichi} simulations is an order of magnitude or two better than that of the {\tt\string NBODY6} simulations. However, we should not only use the overall energy conservation as a measurement of the accuracy or quality of the simulations \citep{dehnen2014fast, wang2020petar}. In fact, for tree and FMM codes, it is a better idea to study the distribution of force errors (section \ref{subsec:accuracy}) along with the evolution of energy to get a better picture \citep{dehnen2014fast}.

\begin{figure*}[ht!]
\gridline{\fig{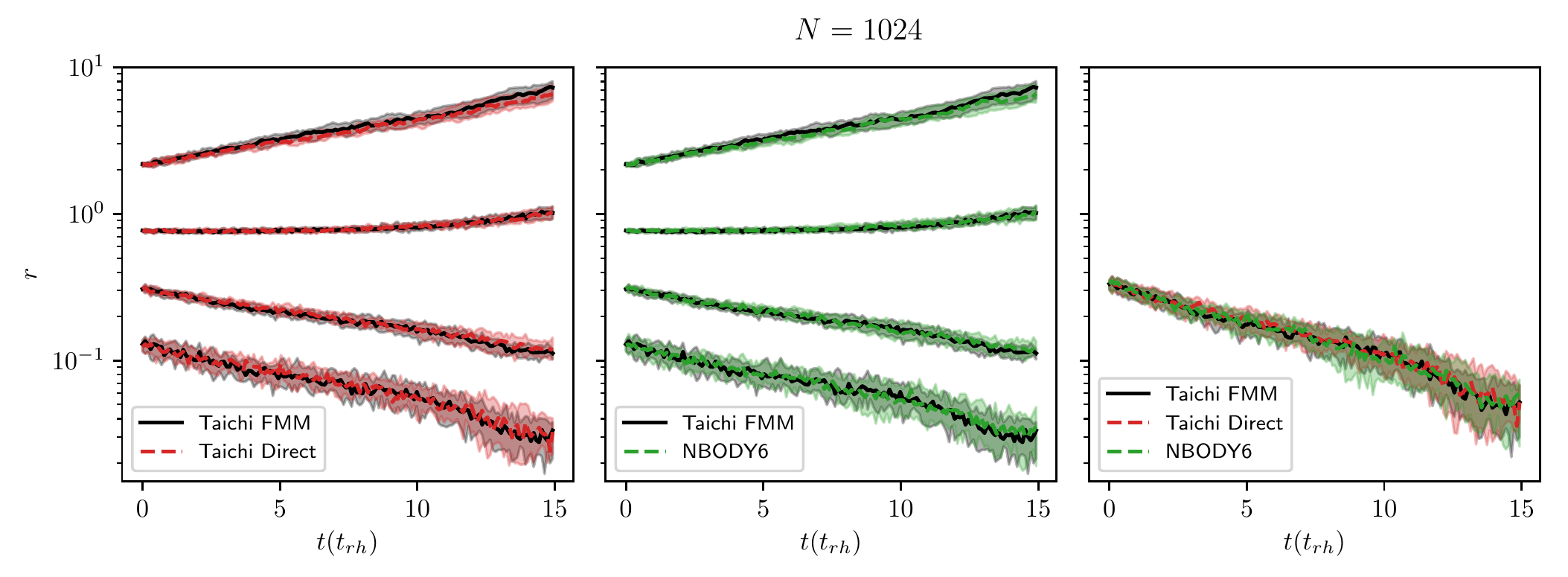}{0.97\textwidth}{}
            }
\gridline{\fig{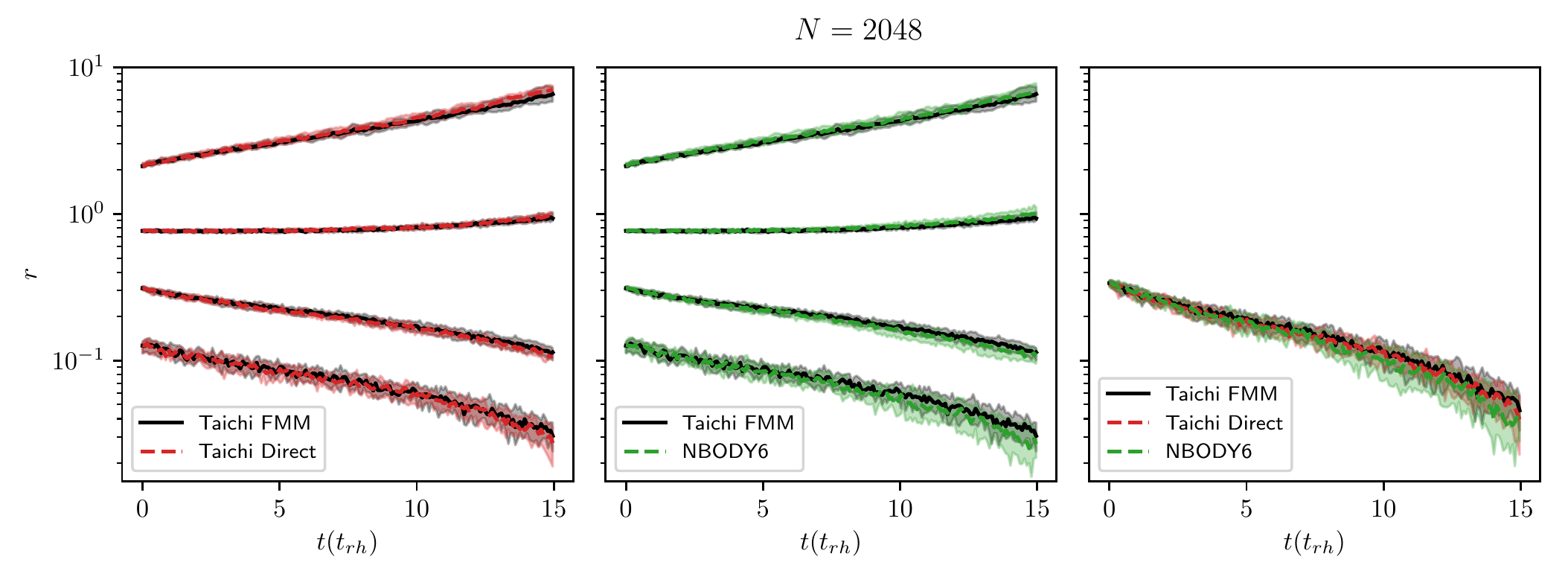}{0.97\textwidth}{}
            }
\gridline{\fig{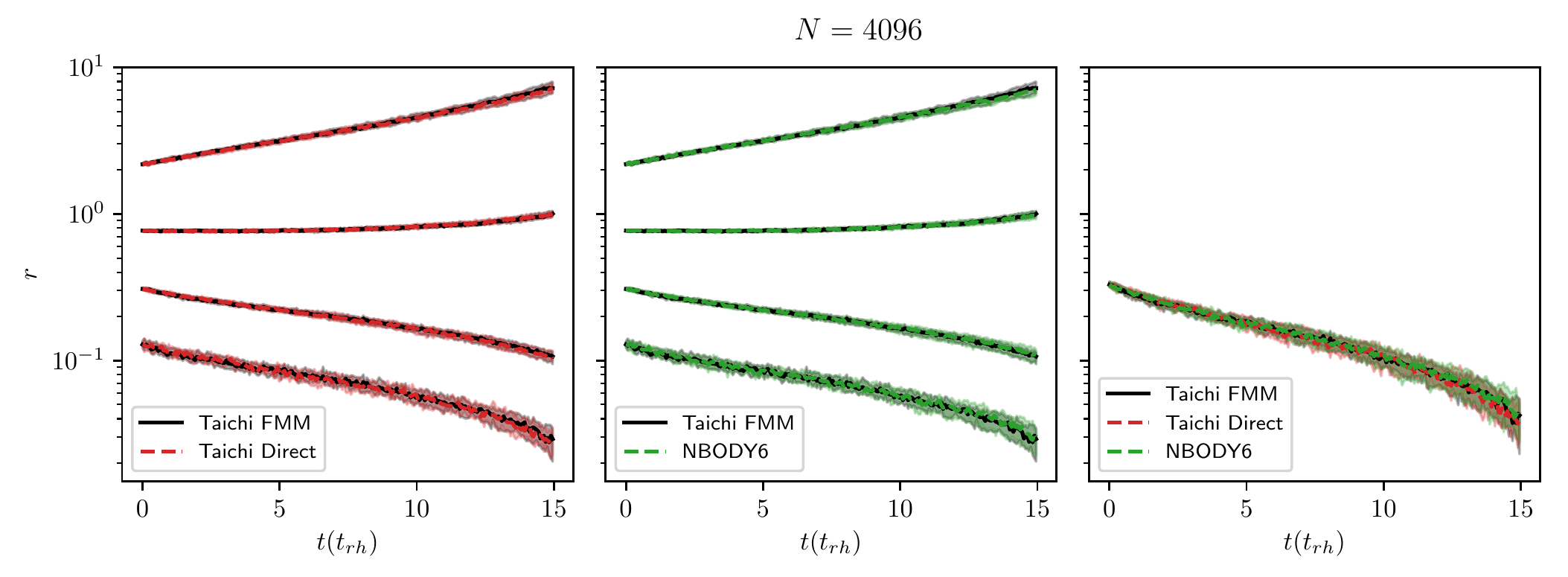}{0.97\textwidth}{}
            }
\caption{Evolution of Lagrangian radius using {\tt\string NBODY6++GPU}
and {\tt\string Taichi} direct and FMM modes. From the bottom to the top, the curves represent 1\%, 10\%, 50\%, and 90\% mass fractions, respectively. The curves have been produced by taking the median of 16 independent realizations in each case. The shaded regions represent the 90th percentile values in each case.  The rightmost plot shows the evolution of the core radius. It can be seen that as $N$ increases, the agreement between FMM and the direct codes gets better.
\label{fig:lagrangian}}
\end{figure*}

\begin{figure*} [ht!]
\gridline{\fig{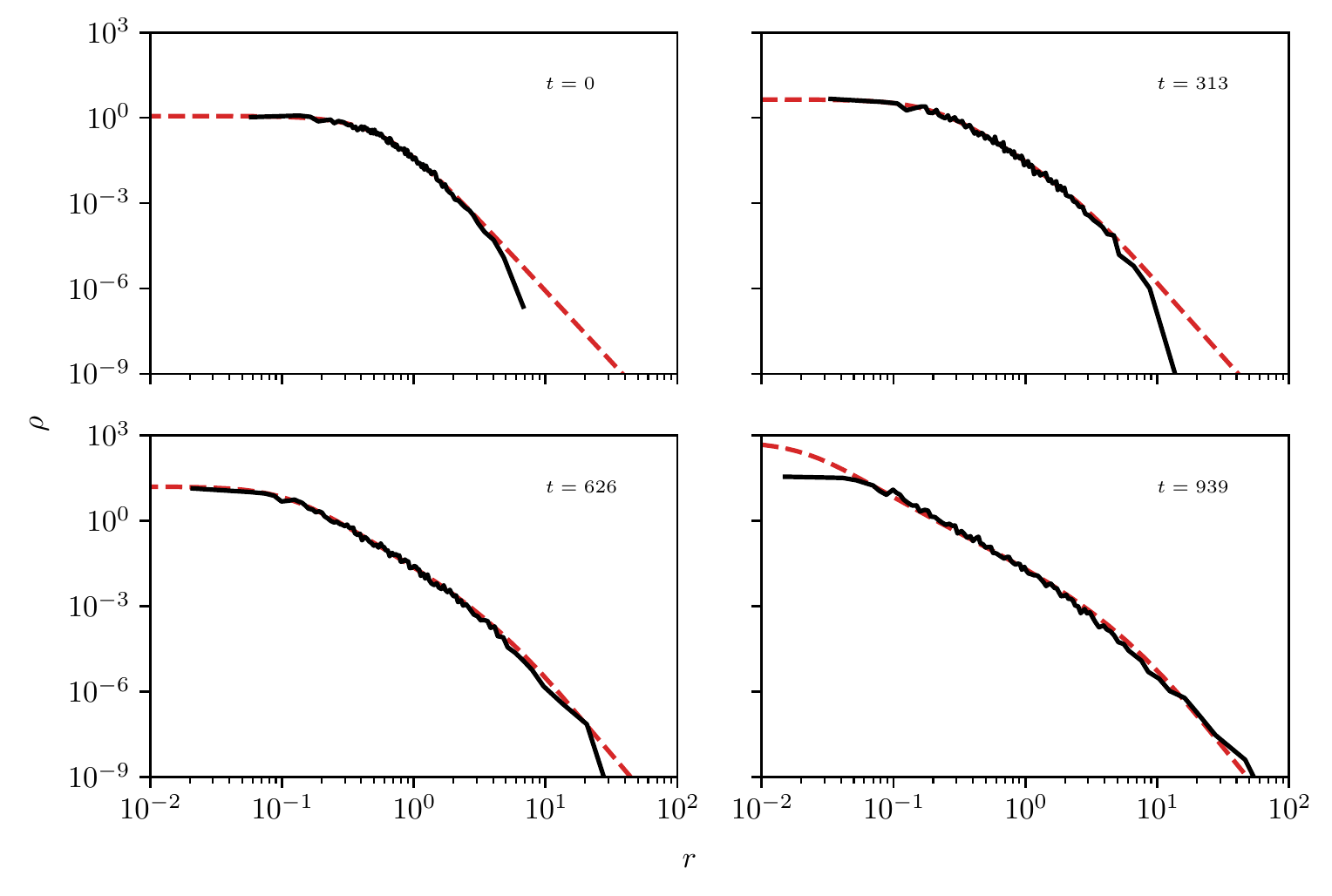}{0.65\textwidth}{}
            }
\gridline{\fig{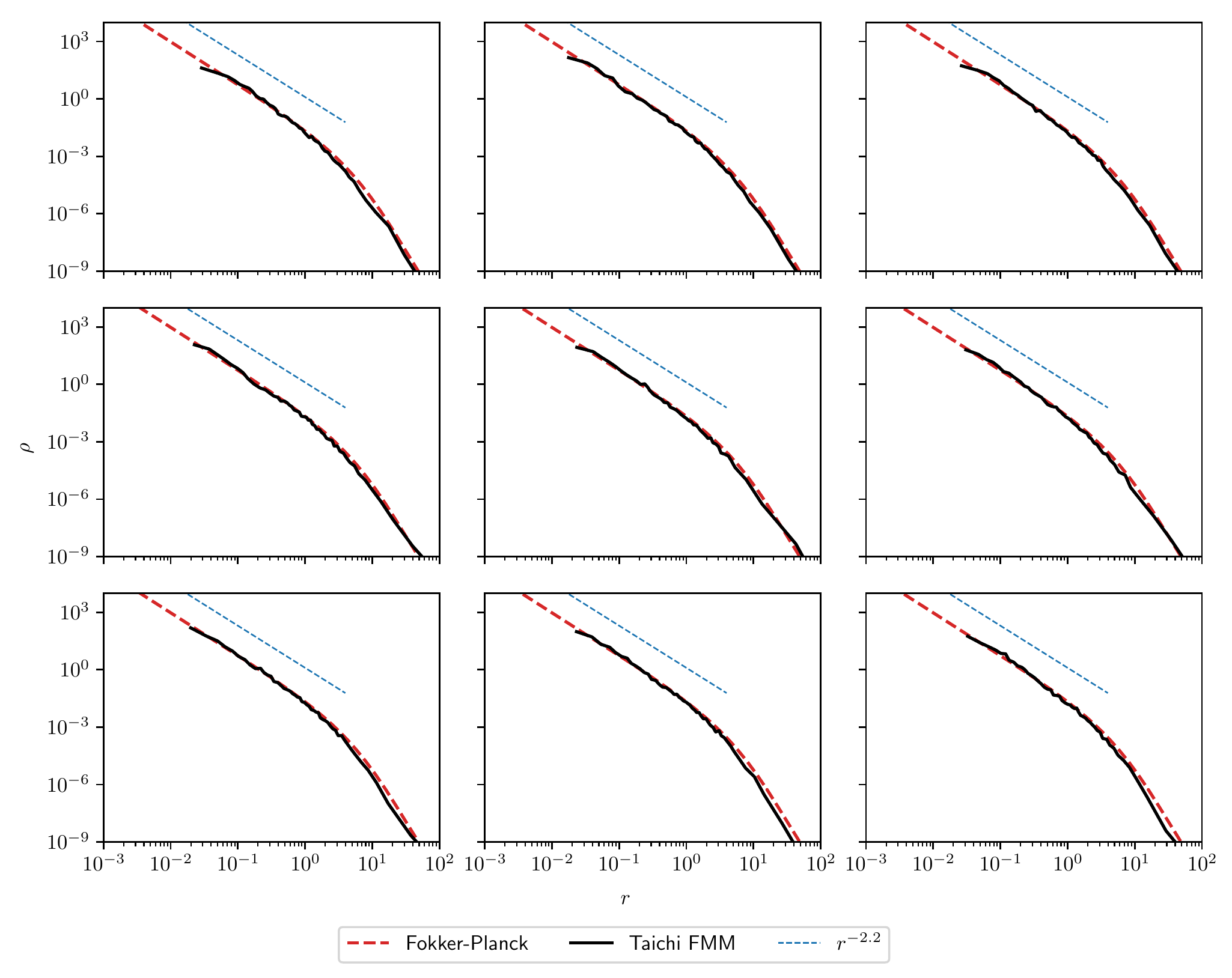}{0.85\textwidth}{}
            }
            
    \caption{The density of the cluster $\rho$ plotted as a function of the radius $r$. Top: the density function of a single 4096-particle realization simulated using FMM is compared to that produced by the Fokker–Planck code at different times during evolution until core collapse. The divergence between the codes at larger radii is caused by the dearth of particles present at larger radii initially.
    Bottom: the density functions of nine independent 4096-particle realizations compared to the density function produced using the Fokker-Planck code at the time of core collpase. The results show significant  agreement between the two codes. This also indicates that the density function agrees with the theoretical power law of the density profile $r^{-2.2}$.  }
    \label{fig:density_func}
\end{figure*}

The first set of tests using the uniform mass Plummer model clusters reveal that {\tt\string Taichi} models the long-term evolution of the clusters properly. This is evident visually from Figure \ref{fig:visual} where we see the formation of the core and from the overlap of the Lagrangian radii curves in Figure \ref{fig:lagrangian}. Here we have utilized {\tt\string AMUSE} \citep{zwart2009multiphysics, pelupessy2013astrophysical, zwart2018astrophysical} to calculate the Lagrangian radius. For the $N=1024$ model simulations, we find that the maximum relative difference between {\tt\string NBODY6} and {\tt\string Taichi} FMM among all 16 realizations is about 4\% for the 1\% mass fraction Lagrangian radius, but the average relative difference is about 1\%. For the half-mass radius, the average relative difference is about 0.3\%. The agreement between {\tt\string Taichi} direct and FMM versions is even better, with the maximum relative difference across all mass fractions being close to $10^{-7}$. As we increase $N$ the agreement between the two methods improves considerably. For example, for the $N=4096$ particle simulations, the maximum relative difference between {\tt\string NBODY6} and FMM in the Lagrangian radii across all four mass fractions is 0.9\%, while the averages range between 0.05\% and 0.1\%.

The core radius is calculated using the definition provided by \cite{casertano1985core}. The core radius is defined as the density-square-weighted sum of the distance from the density center to the particle. Then, the core radius becomes
\begin{equation}
    r_c = \sqrt{\frac{\sum_{i=1}^{N} \rho_{i}^{2} |\mathbf{r_i} - \mathbf{r_d}|^{2} }{\sum_{i=1}^{N} \rho_{i}^{2}}} .
\end{equation}
The density center is defined as
\begin{equation}
    \mathbf{r_{d}} =  \frac{\sum_{i=1}^{N} \rho_{i} \mathbf{r_i}}{\sum_{i=1}^{N} \rho_i}
\end{equation}
where $\rho_{i}$ is the density and is calculated by using a cubic spline kernel over the 32 nearest neighbors from the particle. The density is calculated using {\tt\string HOPInterface} \citep{eisenstein1998hop} present inside {\tt\string AMUSE}.

The core-radius curves in Figure \ref{fig:lagrangian} show agreement between all three codes. For example, in the $N=1024$ particle simulations, the maximum relative difference in core radius is about 0.7\%. With larger $N$ the agreement becomes stronger with smaller deviation between individual simulations. We find that for the $N=4096$ simulations, the maximum relative difference decreases to 0.1\%.

The agreement in the long-term evolution of the Lagrangian and core radii suggests that the evolution of the cluster density should be in agreement. We show this in Figure \ref{fig:density_func}. Comparing the time evolution of the one realization of a 4096-particle model, we find that the density as a function of the radius produced by {\tt\string Taichi} using FMM  matches that of  {\tt\string Phaseflow} at different points during the evolution. Although not presented here, we found a similar picture for simulations with smaller particle numbers. 

What becomes of considerable interest is the behavior of the density function at core collapse. In order to pinpoint the moment of core collpase, we simulate the evolution of an idealized Plummer model until core collapse using {\tt\string Phaseflow} and compare the density functions at the time indicated by {\tt\string Phaseflow} as the core collapse time. We compare nine independent realizations of the 4096 particle model to the idealized density function and find that there is a considerable amount of agreement between them. Some simulations could not be simulated to the core-collapse time owing to the formation of hard binaries (discussed further in section \ref{subsec:integration_issues}). The idealized density profile follows the theoretical density profile $\rho \propto r^{-2.2}$ \citep[e.g.][]{Joshi1999} and thus the density profiles produced by simulating the clusters using FMM also follow the theoretical density profile. This is significant since this phenomenon is purely driven by two-body effects, indicating that {\tt\string Taichi} with FMM can model the two-body relaxation properly, a fundamental aspect of collisional $N$-body simulations.

In order to examine whether there are significant changes with accuracy parameters, we run {\tt\string Taichi} with $\epsilon=10^{-3}$ and $p = 10, 20$. We run a set of five $N=1024$ Plummer model simulations until core collapse to examine the evolution of the Lagrangian radius and compare it to the previous {\tt\string Taichi} FMM results. We also note the energy drift of the simulations using both the softened and unsoftened versions of the code.

\begin{figure*}[ht!]
    \begin{center}
    \includegraphics[width=1.0\textwidth]{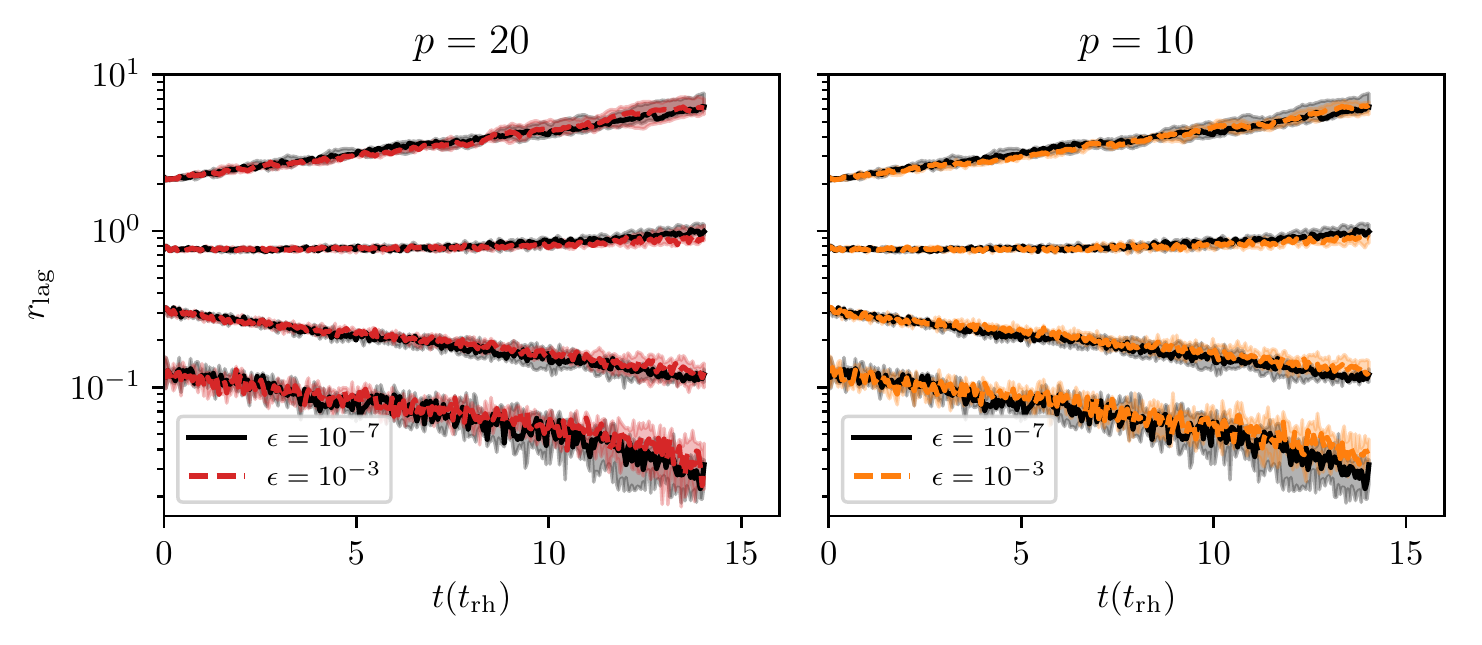}
    \caption{The evolution of the Lagrangian radius for five $N=1024$ simulations is presented here, similar to Figure \ref{fig:lagrangian}. We notice that, even while using lower accuracy parameters, we arrive at a similar evolution of Lagrangian radii of different mass fractions. Compared to the original results, we find that the maximum relative difference in the Lagrangian radii of individual simulations is of the order of $0.001$\%. } 
    \label{fig:lagrangian_low_acc}
     \end{center}
\end{figure*}

From Figure \ref{fig:lagrangian_low_acc}, we find that there is very little change in the overall evolution of the Lagrangian radius compared to that of the more accurate simulations. This suggests that even with lower $\epsilon$ and $p$, {\tt\string Taichi} is properly able to reproduce second-order relaxation effects. This is significant since simulations using lower accuracy parameters are faster. Although not presented here, we find that for the $p=20$ scenario, even though the overall relative energy error does not exceed $10^{-6}$ at any point in time, the relative energy error between any two subsequent snapshots is larger compared to the more accurate simulations. For the $p=10$ scenario, the overall relative energy errors are about $10$x larger compared to the original simulations.

\subsection{Dynamical Friction} \label{subsec:dynamcial_friction}

\begin{figure*}[ht!]
    \plottwo {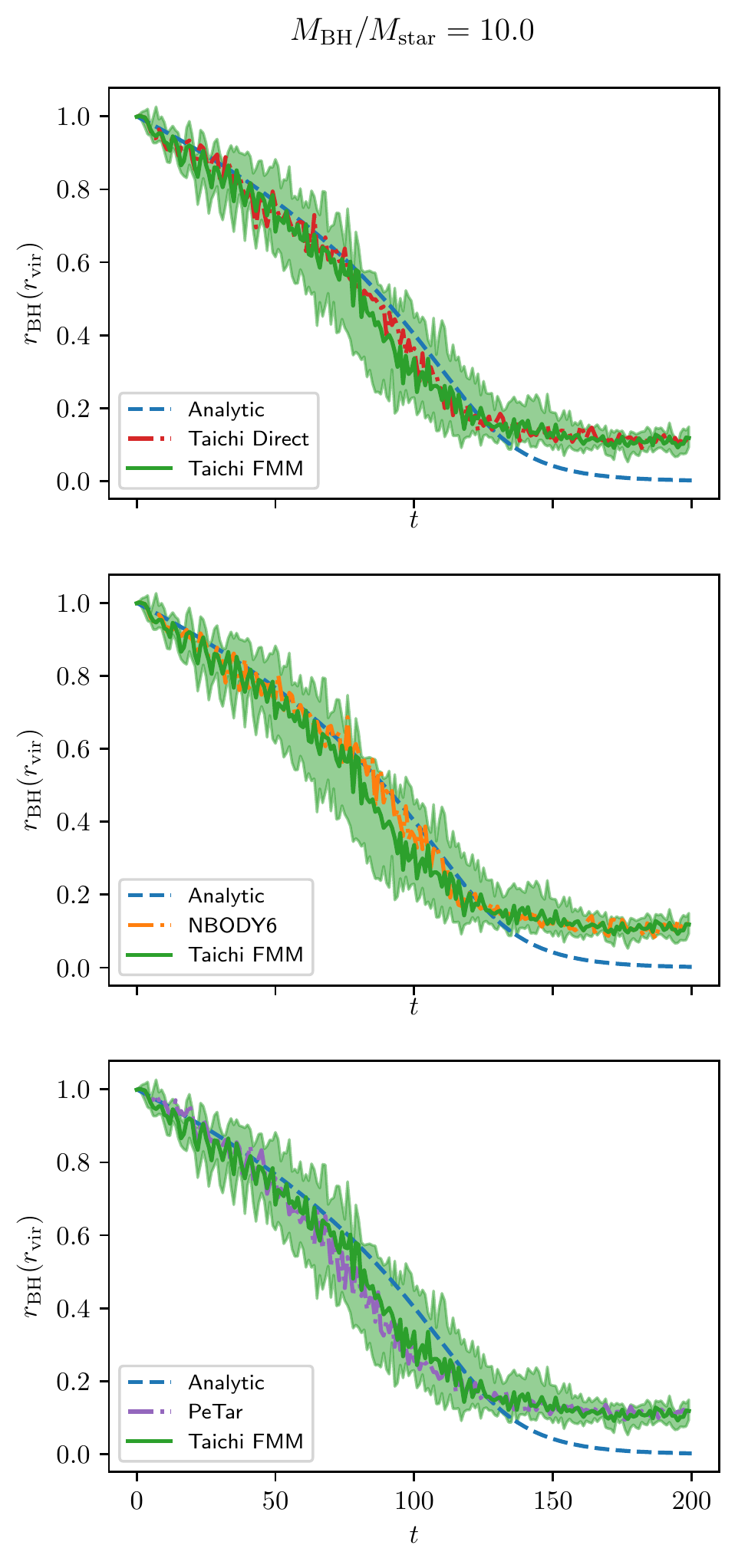} {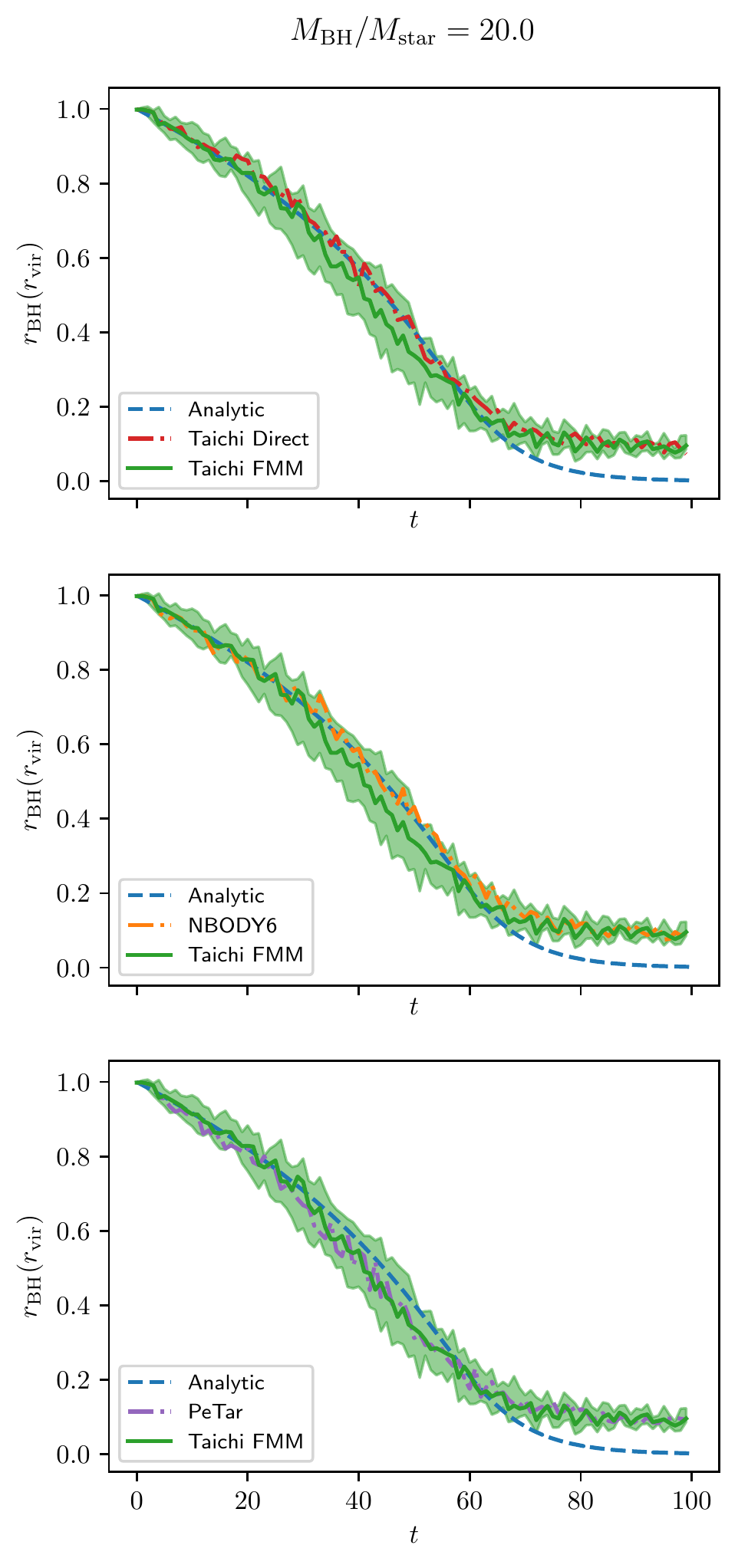}
    \caption{The distance of the massive object ($r_{\mathrm BH}$) is presented as a function of the time (in Henon units) and the virial radius of the cluster. The curves show the inspiral of massive objects of two different masses due to dynamical friction. The solid and dashed curves indicate the median distance of the massive object from the center of mass that was produced after running 30 independent realizations. The shaded regions indicate the 95\% confidence interval values of the median distance for the FMM simulations. All values are binned over one $N$-body time step. We notice that as we increase the mass of the massive object, the agreement between the different methods improves significantly.}
    \label{fig:dynamical_fric_ci}
\end{figure*}

\begin{figure}[ht!]
    \includegraphics[width=0.5\textwidth] {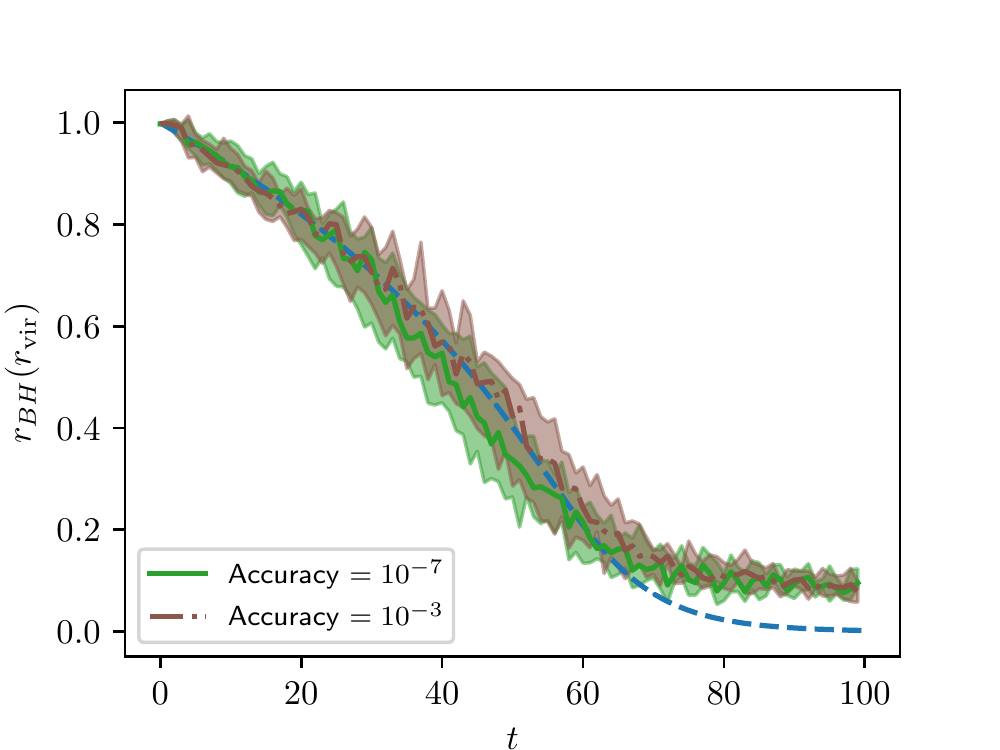}
    \caption{Same as the $M_{\rm BH}/M_{\rm star} = 20.0$ case from Figure \ref{fig:dynamical_fric_ci} but with two different input force accuracies: $\epsilon=10^{-3}$ and $\epsilon=10^{-7}$. Even with an input accuracy four times lower in magnitude compared to the original FMM simulations, the massive particle inspiral time is reproduced very well. }
    \label{fig:dynamical_friction_20_low}
\end{figure}

Another important component of collisional simulations is dynamical friction, which is a purely two-body effect. We can examine it in our tests via the inspiral of a massive object in the field of less massive stars. The rate of inspiral can provide us a direct idea of the ability of a method to reproduce two-body dynamical effects properly. In these tests, we emphasize that we are using the {\tt\string NBODY6} and {\tt\string PeTar} results as a benchmark rather than the analytic results. This is because of the inability of the Chandrasekhar model to reproduce the position of the massive particle near the core. This issue is discussed in detail later.

We notice from figure \ref{fig:dynamical_fric_ci} that {\tt\string Taichi} with FMM is statistically able to reproduce the inspiral rates for both of the tests, agreeing with both the direct $N$-body results and the analytic results. The agreement between individual simulations is, however, not guaranteed. Individual simulations, even though they may agree at the beginning, can vary significantly. For example, for the $M_{\rm BH} / M_{\rm star} = 10.0$ case, the average relative difference between {\tt\string NBODY6} and {\tt\string Taichi} with FMM for the positions of the black hole at the end of the simulation was about 20\%. Even individual simulations performed can vary considerably over multiple runs. The same initial conditions can produce  different inspiral rates if simulated multiple times. It is an artifact of the nonassociativity of floating point operations for multithreaded programs. Even machine-precision errors ($ \sim 10^{-16}$) can grow exponentially over time, and results may diverge after a couple of dynamical times. This is a result of Miller's instability, and this issue has been discussed in more length in a later section. It is thus imperative to perform multiple simulations and use the statistical average of the results rather than results from a single simulation.

The discrepancy between the $N$-body and the analytic results are toward later time steps is caused by the ``core stalling problem." The issue has been noted by \cite{goerdt2006does} and others \citep[e.g.][]{inoue2009test, goerdt2010core} performing $N$-body simulations involving the inspiral of objects in gravitational systems with cores. The stalling represents a flaw in the Chandrasekhar model of relaxation that assumes a Maxwellian distribution of velocity and spherical symmetry that is not perfectly reproduced in discrete models. According to \cite{goerdt2006does}, the stalling is caused by an orbit-scattering resonance in which the perturber and the background reach a stable state. Semianalytic models can be used to correct for the stalling effect \citep[e.g.,][]{petts2016semi,silva2016chandrasekhar}.

The question whether {\tt\string Taichi} FMM using a lower input accuracy and multipole expansion order can reproduce similar inspiral times to that using higher accuracies is interesting. For example, if we used $\epsilon = 10^{-3}$ and $p=10$ instead of $\epsilon = 10^{-7}$ and $p=20$, should we expect results that agree with those from earlier? Figure \ref{fig:dynamical_friction_20_low} suggests that we should in fact find that the results to be in agreement. This is a very important result that suggests that lower-order FMM can be used in cases where we want to model dynamical friction on a few specific particles. 
 Switching to a lower order can save time. In fact, in our simulations, switching to a lower order sped up the simulations by $\sim 2-3$ times. This can have applications in modeling supermassive black hole (SMBH) binaries in a field of smaller stars. We expect the agreement to be good as long as the mass ratio of the massive object to that of the field stars is high enough. Whether low-order FMM can model dynamical friction in cases where the mass ratio is closer to 1 needs to be tested. However, results do suggest that FMM can be used to simulate SMBH binaries and intermediate-mass black hole binaries safely. 

\subsection{Scaling}

\begin{figure*}
    \includegraphics[width=1.0\textwidth]{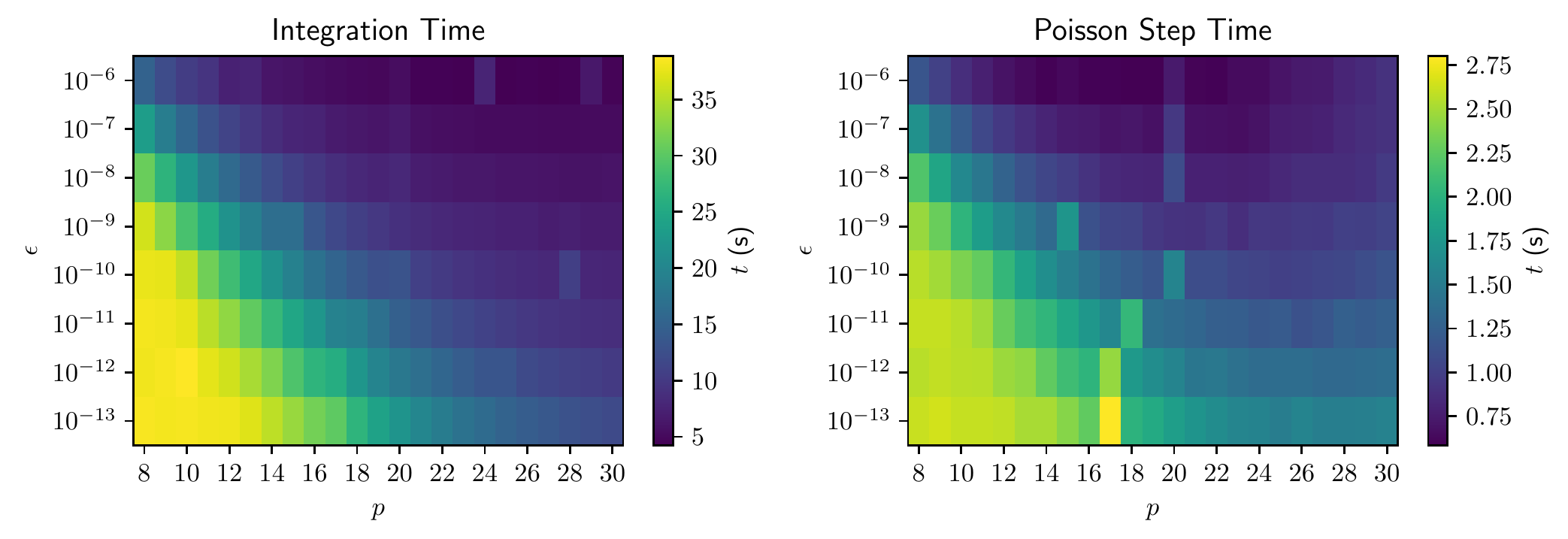}
    \caption{ Heatmaps showing the distribution of wall-clock time as a function of both $p$ and $\epsilon$. Both integration and Poisson step times are determined for evolving a $10^5$ star cluster to 1 timestep. \textit{Left}: The total integration time. It essentially represents how long it takes for {\tt\string Taichi} in total. \textit{Right}: This heatmap only reprsents the amount of time spent computing the forces.}
    \label{fig:timing_all}
\end{figure*}

We perform scaling comparisons and strong scaling tests for {\tt\string Taichi} on a single 28-core {\tt\string Intel Xeon E5-2635 v3} node. All of the wall-clock times have been averaged over wall-clock times from five individual simulations of each realization. The {\tt\string OpenMP} parallelization in Taichi allows us to scale the code over multiple cores in a single node. We run {\tt\string Taichi} FMM with input accuracies of $10^{-7}$ and  $10^{-3}$ and a multipole expansion $p=20$ and compare it to {\tt\string Taichi} direct, {\tt\string NBODY6} and {\tt\string PeTar}. They are then used to perform simulations for $10^3, 10^4, 10^5,$ and $10^6$ particles. The systems are evolved until $t_{\rm final}=1/8192$ $N$-body units. This was used to obtain quick results for large-$N$ systems. A summary of the input parameters has been provided in Table \ref{table:input_par_scaling}.

\begin{table}[h!]

\begin{center}
\begin{tabular}{ |c|c|c| } 
\hline
Code & Input Parameters \\
\hline
\multirow{3}{5em}{{\tt\string Taichi} FMM} & $\epsilon=10^{-7} , 10^{-3}$  \\ 
& $p=20$ \\ 
& $\eta=0.025$ \\ 
\hline
\multirow{2}{5em}{{\tt\string NBODY6}} & {\tt\string NNBOPT}$= 200$  \\ 
& $\eta_{\rm I} = \eta_{\rm R} = 0.01$ \\ 
\hline
\multirow{3}{5em}{{\tt\string PeTar}} & $\theta=0.3$  \\ 
& $\eta_{\rm Hermite}=0.1$ \\
& $\Delta t_{\rm tree} = 1/8192$  \\
\hline
\end{tabular}
\end{center}
\caption{Input parameters used for the scaling tests. Please note that in this case the input parameters for both {\tt\string NBODY6} and {\tt\string PeTar} have been changed slightly compared to Table \ref{table:input_par_summary}.}
\label{table:input_par_scaling}
\end{table}

\begin{figure}[ht!]
    \includegraphics[width=0.48\textwidth]{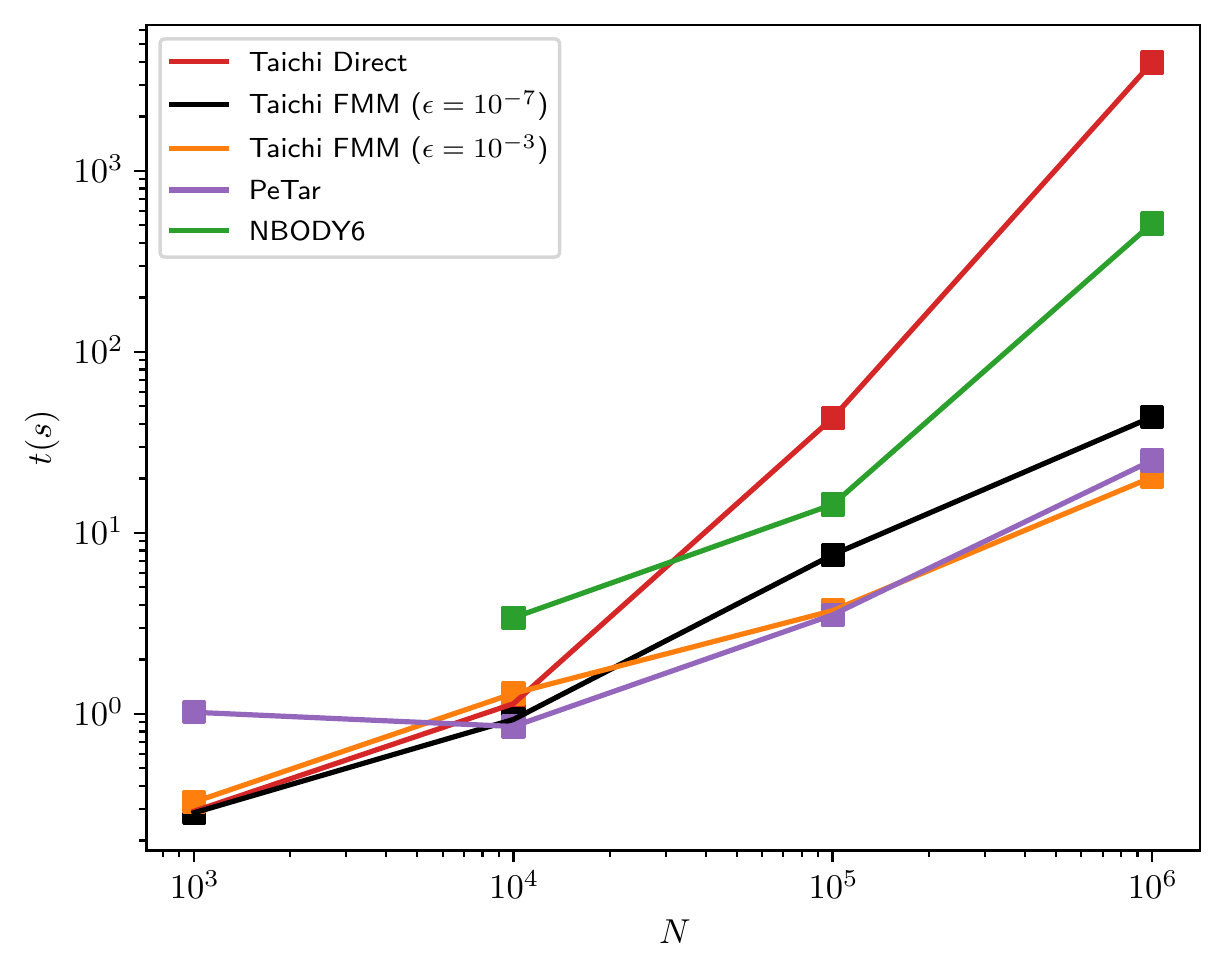}
    \caption{The wall-clock time for one integration step presented as a function of the problem size. For $N<10^{4}$ direct summation is more efficient. However, owing to the $\mathcal{O} (N)$ scaling, for large $N$, FMM becomes highly efficient. }
    \label{fig:scaling_compare}
\end{figure}

To find optimal input parameters, we construct integration and Poisson step time heat maps as shown in Figure \ref{fig:timing_all}. We find that our choice of input parameters ($\epsilon=10^{-7}, p=20$) is optimal. Looking at Figure \ref{fig:scaling_compare}, we notice that FMM is inefficient for simulations with fewer than $\sim 10^4$ particles. This is in part due to the tree building process, which proves to be inefficient compared to the direct algorithm for a  smaller number of particles. However, past that threshold, it becomes more efficient.  For instance, for the million-particle simulation, FMM using $\epsilon = 10^{-7}$ is more than $10$x faster than {\tt\string NBODY6}. We also find that although FMM using $\epsilon = 10^{-7}$ is $1.7$x as slow as {\tt\string PeTar} for a million particles, the version using $\epsilon = 10^{-3}$ is $1.25$x faster than {\tt\string PeTar}. We also find that FMM using smaller $\epsilon$ conserves energy to $1$ part in $10^{12}$ whereas the version using higher $\epsilon$ conserves energy to about $1$ part in $10^{9}$. This is about an order of magnitude smaller than that of {\tt\string PeTar}. 

\begin{figure}[ht!]
    \includegraphics[width=0.5\textwidth]{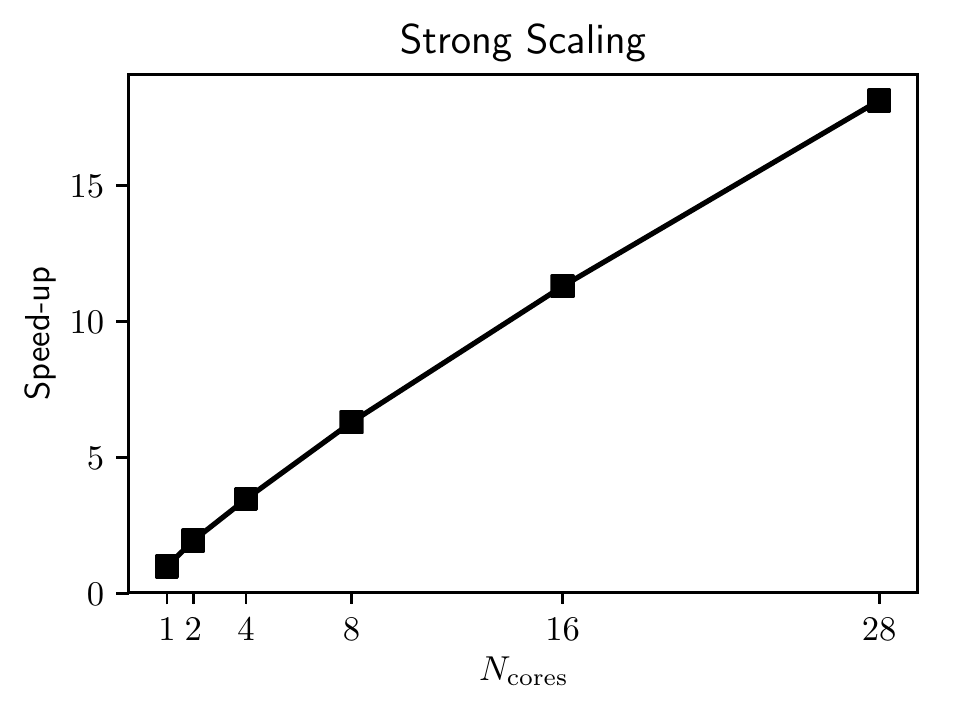}
    \caption{The overall speedup presented as a function of the number of physical cores used. This determines the intranode scaling of the FMM force determination algorithm. The Poisson step time has been used to determine the scaling. The overall scaling follows the same pattern.}
    \label{fig:strong_scaling}
\end{figure}

For the strong scaling test, we simulate a cluster containing $10^6$ stars using $1, 2, 4, 8, 16,$ and $28$ cores. An input accuracy of $10^{-7}$ and a multipole expansion of $p=20$ are used again. The speedup is computed as the ratio of the wall-clock time of the single threaded simulation to that of the multithreaded simulation. As is evident from Figure~\ref{fig:strong_scaling}, {\tt\string Taichi} FMM scales as dictated by Amdahl's law. The graph also indicates that the maximum speedup is not reached on 28 cores and is therefore limited by the number of cores available to us.

\section{Discussion} \label{sec:discussion}

\subsection{Parallelization and Miller's instability} \label{subsec:parallel}
As briefly discussed in section \ref{subsec:dynamcial_friction} floating point arithmetic can play an important part in the outcome of individual simulations. Floating point arithmetic is inherently nonassociative in nature \citep[e.g.][]{villa2009effects}. This is particularly exacerbated in the case of multithreaded floating point operations. For example, reduction operations can lead to different round-off errors during different runs of the same program. In iterative solvers, the results are propagated through various iterations and at the end can produce different round-off errors \citep[e.g.,][]{villa2009effects}. Force calculation relies on iterating over particles and cells and at each time step. As such, any dynamical change in the ordering of threads between two different runs of the same program can lead to discrepancies in the results between two simulations. This is not a feature of the FMM algorithm. This artifact is present in direct summation as well. In our case, analysis reveals that for a simulation containing 1024 particles, the maximum discrepancy between the forces calculated on individual particles between two runs is $\mathcal{O} (10^{-16})$. Out of caution, the serial version of the code was also run multiple times, but no discrepancies were found. This is consistent with round-off errors resulting from dynamical ordering of threads. Any single run of the force algorithm over all particles results in errors of this order. However, even differences of such small order can lead to major discrepancies between the positions and velocities of particles at later times. This is triggered owing to Miller's instability. Over the course of a few dynamical times, the difference between the position and velocity of a particular particle grows exponentially. Although not presented here, we noted that between $t=1$ to $t=10$ for an $N=1024$ particle simulation the maximum difference in the position over all particles grows exponentially from $10^{-16}$ to $10^{-2}$. This is consistent with Miller's instability. This discrepancy only presents itself explicitly when we are looking interested in tracking properties of individual particles. Global properties like energy conservation, evolution of half-mass radius, etc., remain consistent over simulations. This further reiterates the importance of performing multiple simulations and drawing statistical averages rather than relying on  single simulations.

\subsection{Integration Issues} \label{subsec:integration_issues}
Due to a lack of a dedicated regularization scheme, the integrator is sometimes forced to spend a lot of time integrating hard binaries in our test. This ``slowing down" of the simulation becomes more apparent as the simulation approaches core collapse or if they contain primordial binaries. In the process of evolving some of our simulations to core collapse, we noticed that the formation of even one hard binary significantly increased the time required to evolve the system further. For example, in a particular realization containing $1024$ stars, we noticed that the simulation basically halted after 297 time units. Further analysis showed that a binary, with stars having time steps several orders of magnitude smaller than the average time step, was the culprit. One way to alleviate this issue could be to include special treatment of isolated and perturbed binaries such as regularization.

\section{Future Work} \label{sec:future_work}

In a cluster of a million bodies, many primordial binaries are
weakly perturbed for most of the time (or regarded as entirely isolated as in Monte Carlo codes), therefore permitting an 
efficient treatment of their orbits. For an optimal treatment of binary and few-body systems, we seek to integrate a regularization scheme with a future version of {\tt\string Taichi}. 
One of the potential regularization schemes includes Slow Down Algorithmic Regularization \citep[SDAR: ][]{wang2020sdar} which has been included in {\tt\string PeTar}. 

Integration efficiency can also be improved by increasing the integration order, which would allow the usage of larger time steps. A higher-order scheme could also allow the usage of more optimized time step calculation schemes like the Aarseth scheme \citep{aarseth2003gravitational}. Higher-order integration schemes would require the calculation of jerks which is nontrivial with the FMM algorithm. 

We have implemented the approach by \cite{dehnen2014fast} to calculate the jerks, i.e. the time derivative of forces. Therefore, FMM can be incorporated into a traditional fourth-order Hermite codes tat updates the positions and velocities using the information up to jerks. Alternatively, a hierarchical version of the force-gradient integrator recently proposed by \cite[e.g.,][]{Rantala2021} is also promising.

As the Aarseth step function is widely used
in the Hermite integrator, the adaptive stepping is
not time symmetric such that a secular energy
drift is present \citep[e.g.][]{Hut1995, Dehnen2017}.
This energy drift is present even if the time symmetric 
version of \cite{Hut1995} is used. An implicit scheme by
\cite{Makino2006} is proposed, but requires many iterations and is therefore
unpractical. We adopted an approximate time-symmetric method introduced
in \cite{Pelupessy2012}, taking the derivative of time steps
into account. This approach can be generalized with the 
recent method based on the tidal tensor by
\cite{GrudicHopkins2020} as the tidal tensor 
can be easily calculated by FMM, as well as 
its time derivatives.

\section{Conclusion} \label{sec:conclusion}

In this work, we have described a collisional $N$-body code, {\tt\string Taichi}, which incorporates a novel method of calculating forces using FMM. In our implementation, we split up the forces into short range and long range. The former is calculated via direct summation, whereas the latter is calculated using FMM. This results in an algorithmic complexity of $\mathcal O (N)$ rather than the expensive $\mathcal O (N^2)$. This makes post-million-body simulations viable.

Through various tests, we demonstrate that {\tt\string Taichi} can be used to perform collisional stellar system simulations. In the first set of tests, we show that by tuning two input parameters, the mutipole expansion order ($p$) and the input accuracy parameter, ($\epsilon$), we can tightly control the force errors. The median and 99th percentile values are constrained by  the input accuracy. The RMS error values are more weighted toward outliers and can be reduced by increasing $p$ for a given $\epsilon$.

The second set of tests was used to compare long-term behavior of {\tt\string Taichi} with that of {\tt\string NBODY6++GPU}. The relative energy error remained below $10^{-4}$ for {\tt\string Taichi} and only grew sharply as the simulations approached core collapse. The evolution of Lagrangian radii for different mass fractions and core radius shows agreement between both of the codes. This shows that using an approximate force solver like FMM is as good as direct summation for reproducing global properties. Comparison of the density profile with a Fokker-Planck code also shows agreement. At core collapse, the agreement of the density profiles indicates that the realizations simulated with FMM follow the theoretical density power law. This indicates that the approximate force solver is able to reproduce two-body relaxation effects since the theoretical power law is a result of the two-body effects.

Dynamical friction tests allow us to arrive at the same conclusion. The median inspiral times of objects several times the mass of the stars in the clusters closely follow the analytic results and are in agreement with those of {\tt\string NBODY6++GPU}. Furthermore, we demonstrate that we can reproduce proper dynamical friction effects even with a considerably lower input accuracy and multipole expansion order. 

Compared to the direct version of {\tt\string Taichi}, the FMM version speeds up the integration over 100 times for a simulation containing a million stars. We also find that {\tt\string Taichi} FMM is more than $10$x faster than {\tt\string NBODY6++GPU} on a 28-core machine.  However, in our current implementation, FMM becomes effective only for simulations containing more than $10^{4}$ stars. Several bottlenecks are also present in the code. The lack of a proper regularization scheme makes simulations
with binaries virtually impossible. The lack of a higher-order scheme also implies that the code takes smaller time steps which hinders the efficiency. While close binaries or small-$N$ subsystems indeed require extra care, we have shown that approximate force solvers are sufficiently accurate to simulate collective effects due to the uncorrelated two-body encounters in the sense of Chandrasekhar. It is foreseeable that FMM can be combined with Ahmad-Cohen scheme for the regular force calculations.

\acknowledgments

We would like to thank the referee for suggesting valuable changes to improve the paper. We thank Markus Michael Rau for helpful discussions on deriving errors on various statistical averages. We also thank Long Wang for his help with {\tt\string NBODY6++GPU} simulations. In addition, we acknowledge the usage of the Vera cluster, which is supported by the McWilliams Center for Cosmology and Pittsburgh Supercomputing Center, and the Bridges supercomputer which is maintained by the Pittsburgh Supercomputing Center. Q.Z. is supported by the McWilliams Fellowship from the McWilliams Center for Cosmology at Carnegie Mellon University. H.T. acknowledges support from NSF award 2020295.

\vspace{5mm}


\appendix 
\section{Time step symmetrization}

\begin{figure}[ht!]
    \begin{center}            
    \includegraphics[width=0.6\textwidth]{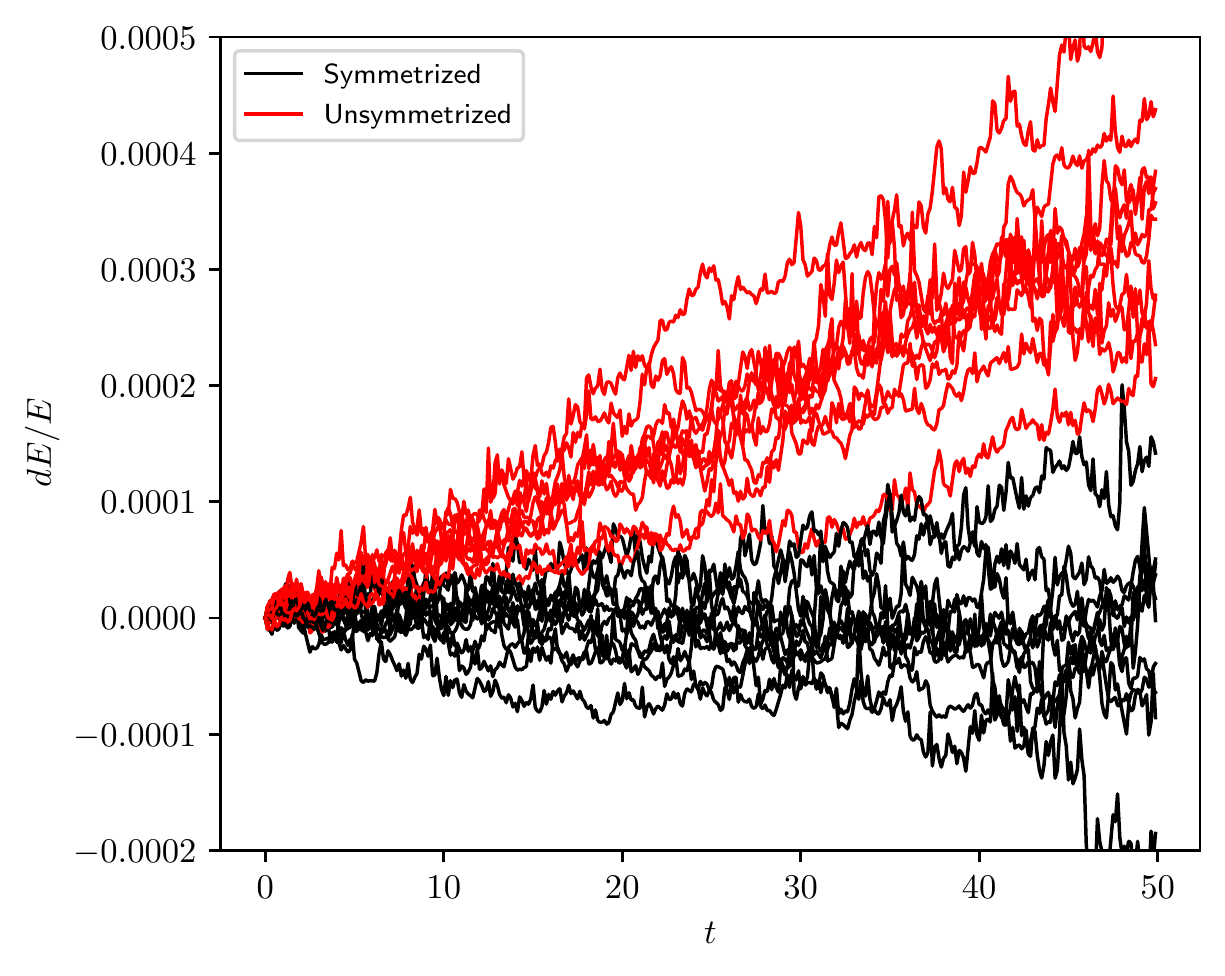}
    \caption{The relative energy error as a function of the $N$-body time here for 10 Plummer model realizations. For this simulation we used the {\tt\string HOLD} integrator and the same softening length and step size as that of \cite{Makino2006}. We find that the symmetrization scheme helps remove the energy drift, identical to \cite{Makino2006} and \cite{Pelupessy2012}.}
    \label{fig:time_symmetrization}    
    \end{center}
\end{figure}

We review the time integration based on the hierarchical 
Hamiltonian splitting here. 
The Hamiltonian of the $N$-body system consists of the
total potential term 
\begin{equation}
\mathcal{V} = \sum_{i<j} \sum_j \frac{m_i m_j}{r_{ij}} ,   
\end{equation}
and the total kinetic term
\begin{equation}
\mathcal{T} = \frac{1}{2}\sum_{i} \frac{\mathbf{p}_{i}^2}{m_i}.
\end{equation}
For most collisional systems, 
there exists a wide
range of dynamical timescales, defined by both the smooth orbit of a particle in the mean field potential and 
interactions between individual particles. To speed up the calculations, individual
time stepping has been used since
\cite{aarseth2003gravitational}. In \cite{Pelupessy2012},
this was implemented by splitting the Hamiltonian
\begin{equation}
\mathcal{H} = \mathcal{H}_{F} + \mathcal{H}_{S},
\end{equation}
where the original Hamiltonian 
$\mathcal{H} =\mathcal{V} + \mathcal{T}$ is decomposed 
into a fast and slow subsystem based on the step size
assigned to each particle. We adopt the step size
criteria from \cite{Pelupessy2012}, which is constrained 
by both the freefall time,
\begin{equation}
\tau_{\rm freefall} = \eta \sqrt{\frac{r_{ij}} {a_{ij}}}
\label{eq:step_freefall}
\end{equation}
and a flyby time 
\begin{equation}
\tau_{\rm flyby} = \eta \frac{r_{ij}} {v_{ij}}.
\label{eq:step_flyby}
\end{equation}
The slow system contains 
the contributions from both slow particles and 
the cross interaction between slow
and fast particles as
\begin{equation}
\mathcal{H}_{S} =  \mathcal{T}_{S} + \mathcal{V}_{SS} +  \mathcal{V}_{FS},
\end{equation}
The fast subsystem now only consists of fast particles
\begin{equation}
\mathcal{H}_{F} =  \mathcal{T}_{F} + \mathcal{V}_{FF},
\end{equation}
where $\mathcal{T}_{F}$ is the kinetic energy of fast
particles and $\mathcal{V}_{FF}$ consists of potential 
energy solely from fast particles.
Now, $\mathcal{H}_{F}$ can be integrated separately 
from the slow system, 
where the forces between the fast and slow systems need to be
calculated at the pace of the slow system. The integration
then proceeds recursively to $\mathcal{H}_{F}$.
The above procedure leads to a second-order accurate and
momentum-conserving scheme. One subtle point is that the use
of individual time steps breaks the time symmetry
of the leap-frog integrator, leading to a drift in the total energy. 
To counter this, we adopt an approximate time-symmetric
stepping function that is introduced by 
\cite{Pelupessy2012} which removes the iterations 
required by an implicit time-symmetric scheme 
\citep{Hut1995, Makino2006}. The idea is to incorporate 
the time derivatives of Eq.~(\ref{eq:step_freefall}) 
and ~(\ref{eq:step_flyby}) to construct a first-order-accurate estimate of step size into the future according to
\begin{equation}
\tau_{\rm sym} = {\tau} {(1 - \frac{1}{2}\frac{d \tau}{dt}) ^{-1}}
\label{eq:step_symm}
\end{equation}
As a result, this treatment 
removes a secular energy drift often associated 
with individual time steps in long-term evolution of
$N$-body systems. An alternative form to 
Eq.~\ref{eq:step_symm} is given by \cite{Rantala2021}.
In Figure~\ref{fig:time_symmetrization}, we simulate 10 
different $N=100$ Plummer models to 50 $N$-body units and find that the time-symmetrized version of the {\tt\string HOLD} 
integrator conserves energy better than the unsymmetrized version.

\section{Energy error associated with hard binaries}
{\tt\string NBODY6} has sophisticated treatment of binaries
and multiples, which are absent in {\tt\string Taichi}. To access the 
impact of hard binaries, we modify the Newtonian gravity for both {\tt\string NBODY6} and {\tt\string Taichi} to incorporate a Plummer softening $\phi(r) = (r^2+\epsilon_{\rm soft}^2)^{-1/2}$, where $\epsilon_{\rm soft}$ is the softening length.  We then
simulate Plummer models of $N=1024, 2048$, and $4096$ with $\epsilon_{\rm soft} = \frac{1}{N}$. This estimate of close encounter distance follows from \cite{dehnen2011n}, where $\epsilon_{\rm 2body} = \frac{2R}{N}$ where $R$ is a characteristic radius of the system. In $N$-body units, $2R \sim 1$, so $\epsilon \sim \frac{1}{N}$.  

\begin{figure*}[ht!]  
    \begin{center}
    \includegraphics[width=1.0\textwidth]{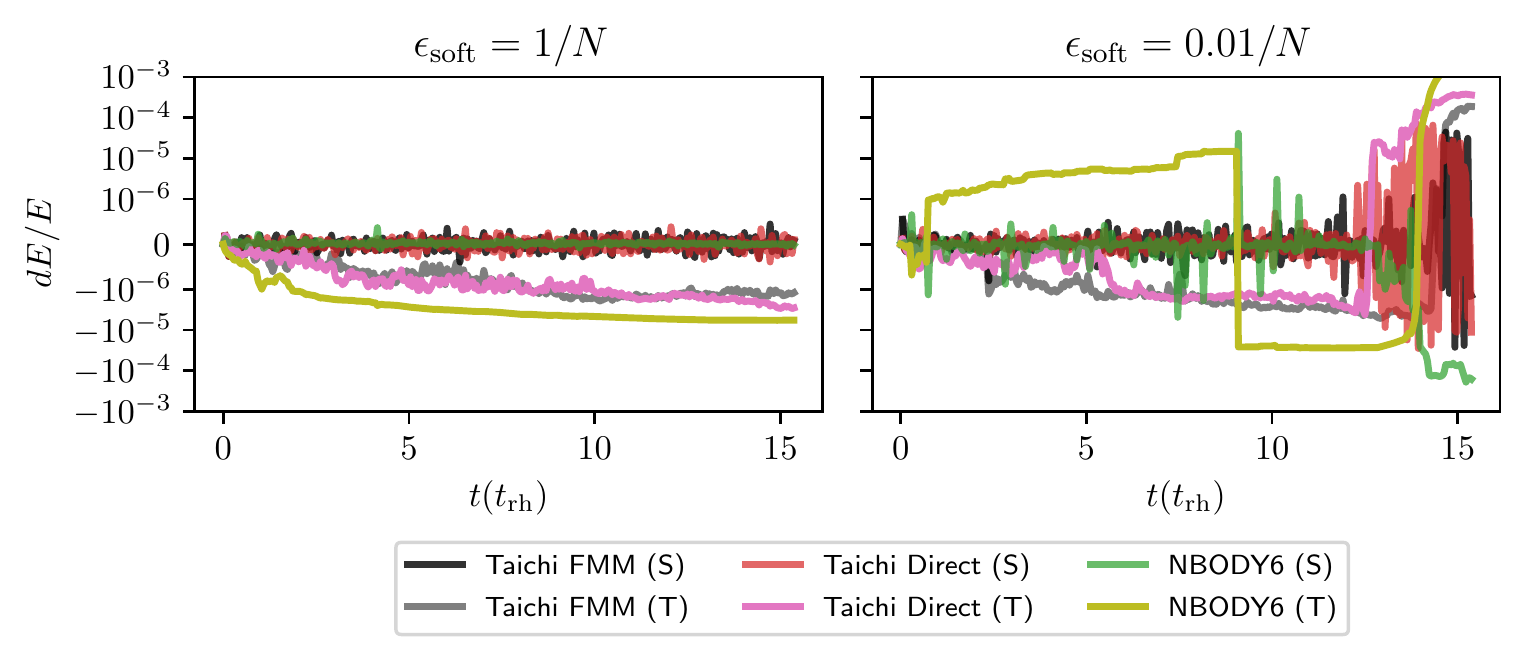}
    \caption{ The relative energy error is presented as a function of the relaxation time for the evolution of a single $N=1024$ Plummer model realization. The softening lengths used in this simulation are $1/N$ and $0.01/N$. The energy drift in the former case is smooth, which is what is expected when force softening is used. We find that the energy drift is virtually similar for both {\tt\string Taichi} direct and FMM and is about an order of magnitude or two better than that of {\tt\string NBODY6}. We find that for the latter simulation, which uses a lower softening, there are jumps in energy that are caused by close encounters.}
    \label{fig:energy_cons_soft_compare}
    \end{center}    
\end{figure*}

\begin{figure*}[ht!]
    \begin{center}
    \includegraphics[width=1.0\textwidth]{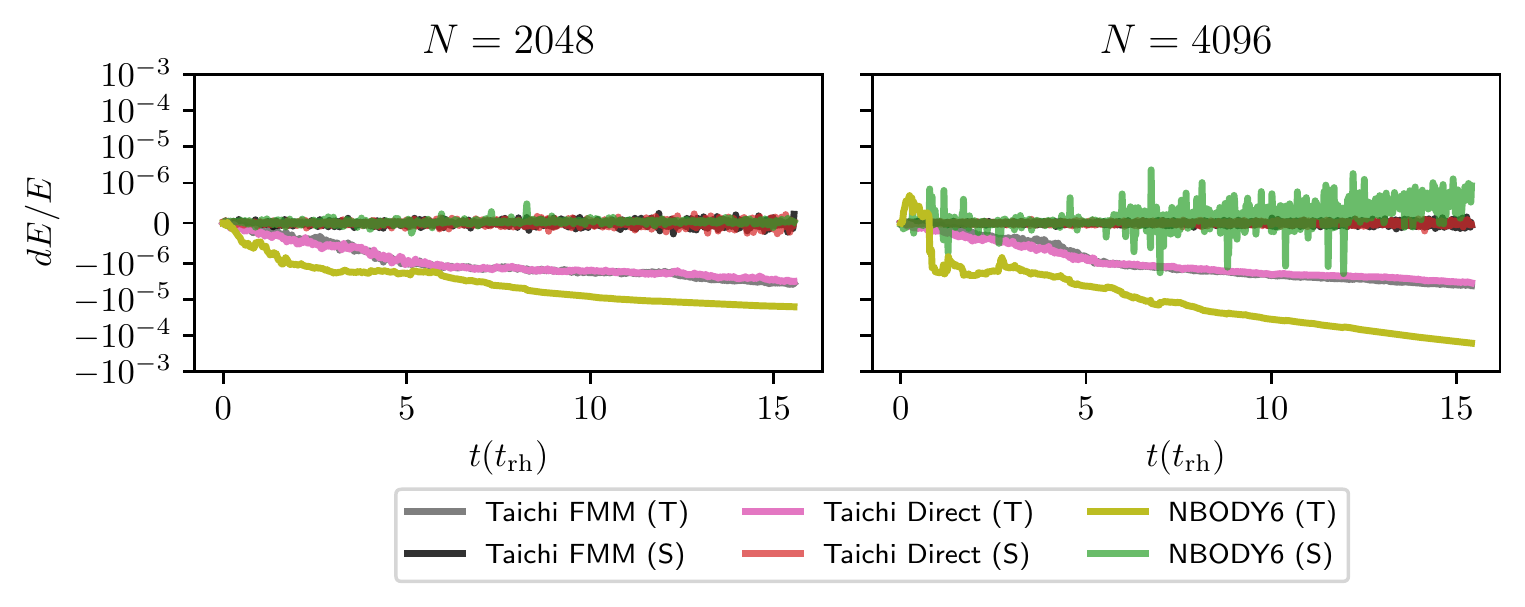}
    \caption{ The relative energy error is presented as a function of the relaxation time for the evolution of single $N = 2048$ and $4096$ Plummer models. The softening length used in this simulation is $1/N$. The energy drift is smooth, which is what is expected when force softening is used. This is similar to what we found in Figure \ref{fig:energy_cons_soft_compare}.}   
    \label{fig:e_cons_eps}
     \end{center}
\end{figure*}

We find that when softening is enabled, the drift in energy is smooth over time and is virtually similar for both {\tt\string Taichi} direct and FMM. We notice this for all simulations as shown in Figures~\ref{fig:energy_cons_soft_compare} \& \ref{fig:e_cons_eps}. This signals that the jumps in energy observed before are primarily related to close encounters. In the absence of close encounters, FMM does not lead to any systemic bias in the energy drift compared to its direct counterpart. In case of {\tt\string NBODY6} the relative energy error for the softened version is an order of magnitude or two times larger than that of {\tt\string Taichi}. We speculate that this could be because of the lack of symmetrized time steps or because the code is not optimized to run with softening enabled. 

We also run a separate $N=1024$ particle simulation with $\epsilon_{\rm soft} = \frac{0.01}{N}$. A smaller softening would allow closer encounters: hence, we should expect jumps in energy similar to the unsoftened versions of the codes. This is indeed what we have observed. The impact of different softening is is also reported in \cite{MaureiraFredes2018}, in a fourth-order Hermite $N$-body code with Plummer softening.  Our findings are  consistent with \cite{MaureiraFredes2018}. \cite{Nitadori2008} have conducted a  comparison between fourth-, sixth- and eighth-order Hermite schemes. The energy error jumps are absent in the eigth-order run, primarily due to shortened step size in the \textit{outer} region. Therefore, we suspect that the cause of energy error jumps is a mismatch between those high-speed particles ejected after close encounters and those particles assigned with large step size (i.e., in the outer region). This mirrors a well-known issue with SPH using individual time steps \citep{Saitoh2009}. Interestingly, \cite{Lockmann2008} adapted a strategy of detecting fast approaches and close encounters to prevent sudden changes in their Hermite integrator. More studies are needed to verify this assertion. 

\section{Dynamical Friction}

\begin{figure*}[ht!]
    \plottwo {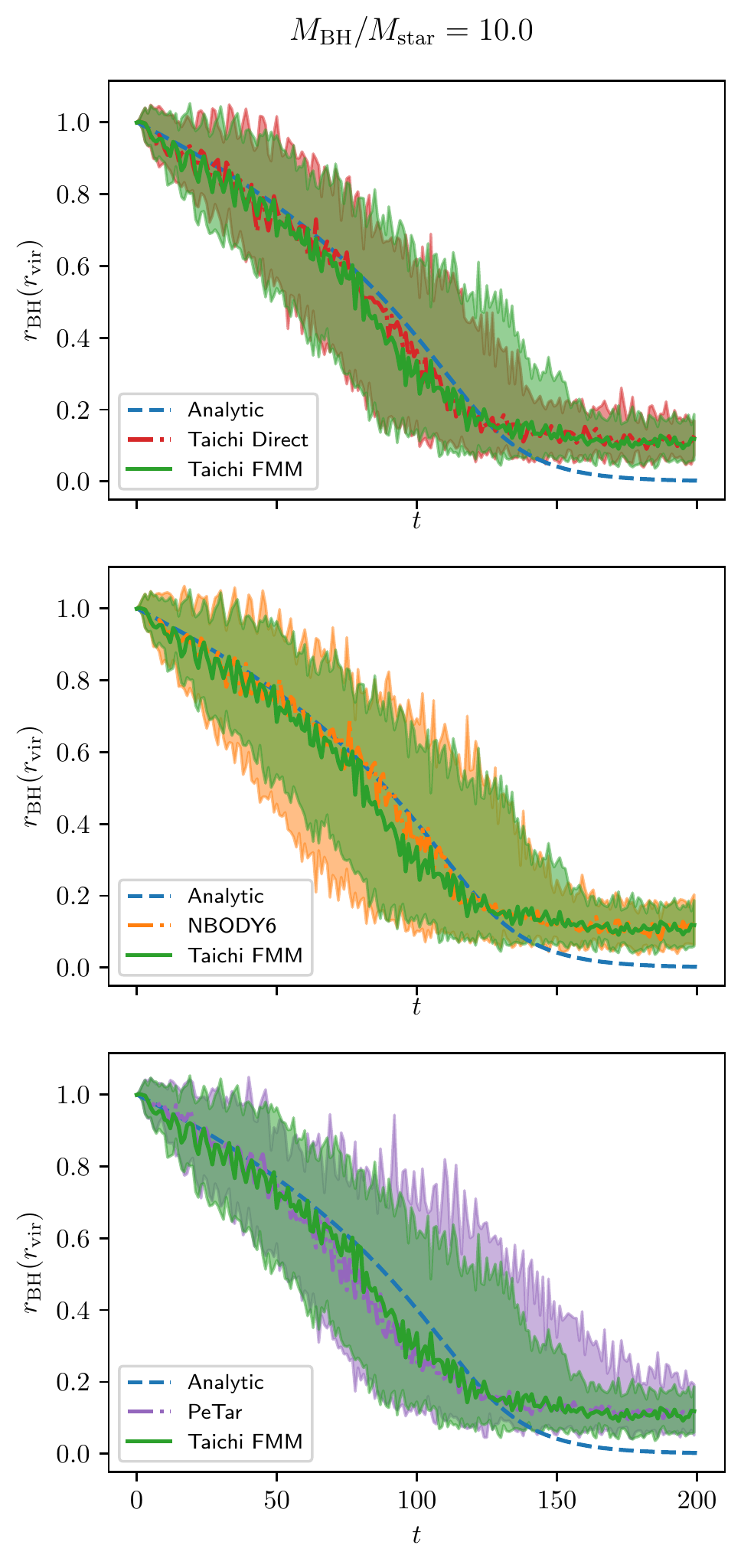} {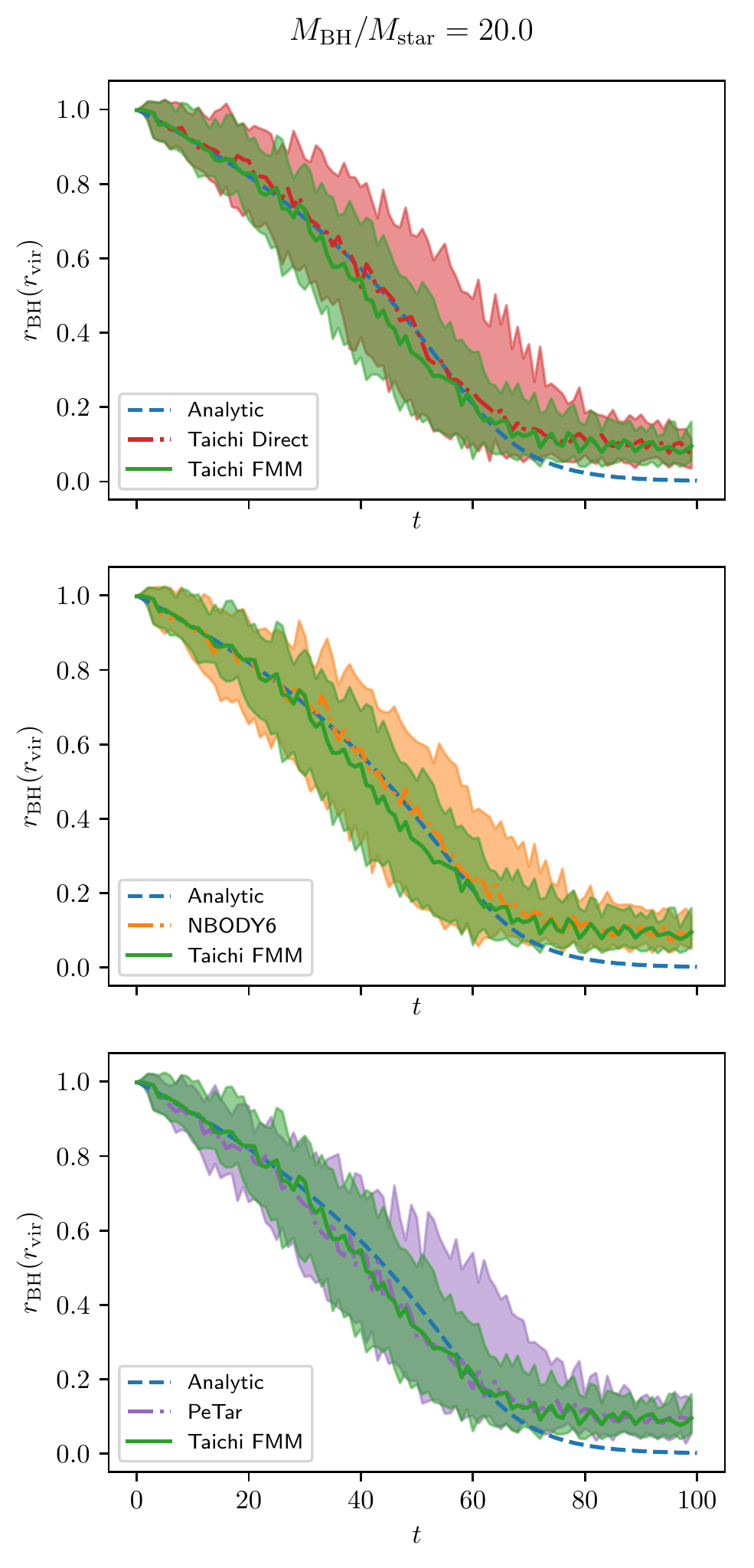}
    \caption{The median distance of the massive object is presented as a function of the time (in Henon units) and the virial radius of the cluster. Unlike Figure \ref{fig:dynamical_fric_ci}, the shaded regions in this figure indicate the spread of radius of the black hole particle from the center of mass of the cluster. Presented here are the 90th percentile values of the distance. All values are binned over one $N$-body time step. One can see the large spread of radii, indicating the inherent stochasticity present in the simulation.  }
    \label{fig:dynamical_fric}
\end{figure*}

One important point that we noted from the dynamical friction test was that there was a lot of stochasticity involved with the position of the black hole particle. This is especially prominent in the case of the less massive black hole. The stochastic nature of the inspiral has been observed and noted in \cite{rodriguez2018new} as well. To present an idea of how the spread of the positions of the massive particle vary with different masses and methods, we present Figure \ref{fig:dynamical_fric}. We notice that as the mass increases, the spread of the positions becomes smaller. This indicates that the $M_{\rm BH} / M_{\rm star} = 10.0$ case is more sensitive to force discrepancies and round-off errors. Indeed, we noticed that the discrepancy between direct-summation-based $N$-body codes and codes using approximate solvers is more noticeable for that case. Increasing the mass, however, reduces the difference.

\end{document}